\documentclass[]{aa}  
\usepackage{graphicx}
\usepackage{txfonts}
\usepackage{lipsum}
\usepackage[dvipsnames]{xcolor}
\usepackage{stfloats}
\usepackage{multirow}
\usepackage[version=4]{mhchem}

\usepackage{array}
\newcolumntype{L}[1]{>{\raggedright\let\newline\\\arraybackslash\hspace{0pt}}m{#1}}
\newcolumntype{C}[1]{>{\centering\let\newline\\\arraybackslash\hspace{0pt}}m{#1}}
\newcolumntype{R}[1]{>{\raggedleft\let\newline\\\arraybackslash\hspace{0pt}}m{#1}}
\newcommand\Bstrut{\rule[-1.5ex]{0pt}{0pt}}   % = `bottom' strut
\newcommand*{\dprime}{^{\prime\prime}\mkern-1.2mu}

\begin{document} 
   \title{Disentangling protoplanetary disk gas mass and carbon depletion through combined atomic and molecular tracers}
   \titlerunning{Disentangling protoplanetary disk gas mass and carbon depletion through combined tracers}

   \author{J.A. Sturm\inst{1}
          \and
          A.S. Booth\inst{1}
          \and
          M.K. McClure\inst{1}
          \and
          M. Leemker\inst{1}
          \and
          E.F. van Dishoeck\inst{1,2}
          }

   \institute{
            Leiden Observatory, Leiden University, P.O. Box 9513, NL-2300 RA Leiden, The Netherlands, e-mail: sturm@strw.leidenuniv.nl
            \and
            European Southern Observatory, Karl-Schwarzschild-Strasse 2, 85748 Garching bei M\"unchen, Germany
            }
    \date{Received XXX; accepted YYY}

    \abstract
   {The total disk gas mass and elemental C, N, O composition of protoplanetary disks are crucial ingredients for our understanding of planet formation. Measuring the gas mass is complicated, since H$_2$ cannot be detected in the cold bulk of the disk and the elemental abundances with respect to hydrogen are degenerate with gas mass in all disk models.}
   {We aim to determine the gas mass and elemental abundances ratios C/H and O/H in the transition disk around LkCa~15, one of the few disks for which HD data are available, combining as many chemical tracers as possible.}
   {We present new NOrthern Extended Millimeter Array observations of CO, $^{13}$CO, C$^{18}$O and optically thin C$^{17}$O $J$=2-1 lines, and use additional high angular resolution Atacama Large Millimeter Array  millimeter continuum and CO data to construct a representative model of LkCa~15.
   Using a grid of 60 azimuthally symmetric thermo-chemical Dust And LInes disk models, we translate the observed fluxes to elemental abundances and constrain the best fitting parameter space of the disk gas mass.}
   {The transitions that constrain the gas mass and carbon abundance most are C$^{17}$O 2-1, N${_2}$H$^+$ 3-2 and HD 1-0. Using these three molecules we find that the gas mass in the LkCa~15 disk is $M_\mathrm{g}=0.01 ^{+0.01}_{-0.004}~M_{\odot}$, a factor of six lower than estimated before. This value is consistent with cosmic ray ionization rates between $10^{-16} - 10^{-18}$ s$^{-1}$, where $10^{-18}$ s$^{-1}$ is a lower limit based on the HD upper limit. The carbon abundance is C/H~=~($3 \pm 1.5) \times10^{-5}$, implying a moderate depletion of elemental carbon by a factor of 3-9. All other analyzed transitions also agree with these numbers, within a modeling uncertainty of a factor of two. Using the resolved \ce{C2H} image we find a C/O ratio of $\sim$1, which is consistent with literature values of  H$_2$O depletion in this disk.
   The lack of severe carbon depletion in the LkCa~15 disk is consistent with the young age of the disk, but contrasts with the higher depletions seen in older cold transition disks.}
   {Combining optically thin CO isotopologue lines with N$_2$H$^+$ is promising to break the degeneracy between gas mass and CO abundance. The moderate level of depletion for this source with a cold, but young disk, suggests that long carbon transformation timescales contribute to the evolutionary trend seen in the level of carbon depletion among disk populations, rather than evolving temperature effects and presence of dust traps alone. HD observations remain important to determine the disk gas mass.}

    \maketitle
%-------------------------------------------------------------------
%\vspace*{-1 mm}\
\section{Introduction}
\label{sec:intro}
The total mass of a protoplanetary disk is one of its most important quantities. 
The disk mass sets the potential for planet formation at different evolutionary stages, and it is a fundamental input for models of the disk chemical inventory, planet formation, and planet-disk interactions \citep[see amongst others][]{2007prpl.conf..591L,2012A&A...541A..97M}.
In contrast to the dust mass, which can be relatively well constrained by the spectral energy distribution and resolved mm continuum observations \citep{1983QJRAS..24..267H}, the gas mass is much harder to determine.
Over 99\% of the gas resides in \ce{H_2} and He, which cannot be directly observed in the bulk of the disk due to lack of a dipole moment.
Hydrogen deuteride (HD) emits from the warmer atmospheric layers in the disk and can be used to determine the gas mass within a factor of a few \citep{2013ApJ...776L..38F,2016ApJ...831..167M,2017A&A...605A..69T}.
However, only a handful of the most massive sources have a detection of the HD 1-0 line, using far-infrared \textit{Herschel} observations \citep{2013Natur.493..644B,2016ApJ...831..167M}, and, unless a far-infrared facility is selected for NASA's upcoming probe class mission, no instruments will be able to target HD in the near future.
For several massive disks, \textit{Herschel} HD upper limits can provide additional constraints on the gas mass \citep{2020A&A...634A..88K}.

Carbon monoxide (CO), the second most abundant molecule in disks, is often used to trace the disk mass using a scaling factor to find the total gas mass.
Early studies of a few disks find unexpectedly weak CO emission \citep{2001A&A...377..566V,2003A&A...402.1003D,2008A&A...488..565C,2013ApJ...776L..38F}.
Recent large surveys in the nearby low-mass star-forming regions Lupus and Chamaeleon \citep[][]{2016ApJ...828...46A,2018ApJ...859...21A,2017ApJ...844...99L,2017A&A...599A.113M} suggest that weak CO emission is common, and that CO might be 1-2 orders of magnitude less abundant in protoplanetary disks than the canonical ISM value of C/H = 1.35$\times$10$^{-4}$.
This depletion is greater than the amount of depletion expected purely from freezeout and photo-dissociation.
The CO is thought to be in ices on large dust grains in the midplane and out of reach of chemical equilibrium, which could result in low elemental C/H and O/H abundances in the gas compared to the ISM value \citep[see e.g.,][]{2016A&A...592A..83K,2022arXiv220104089S}. 
These elemental ratios are very important for the disk chemistry, and understanding the mechanisms that set these values is crucial to understand the formation of habitable planets.

Assessing the degree of CO depletion is complicated by several factors. 
First, CO depletion is thought to be a combination of dust grain growth, radial and vertical dust dynamics, CO freezeout, and chemical processing of CO \citep{2001A&A...377..566V, 2014ApJ...788...59W,2017A&A...599A.113M, 2017MNRAS.469.3994B, 2018ApJ...856...85S, 2018A&A...618A.182B, 2018ApJ...864...78K, 2019ApJ...883...98Z, 2020ApJ...899..134K}.
Second, recent studies that observed the rare isotopologues \ce{^{13}C^{18}O} \citep{2020ApJ...891L..16Z} and \ce{^{13}C^{17}O} \citep{2019ApJ...882L..31B} show that even \ce{C^{18}O} lines can be optically thick in the densest, inner regions of the disk, attributed to enhanced abundances due to radial drift of ice rich material on large dust grains within the snowline.
Third, the C/O ratio in the gas is thought to be radially varying between 1 at large radii and $\sim$0.47 at small radii because of the freezeout of key C and O bearing species like \ce{H_2O}, \ce{CO} and \ce{CO_2} \citep{2011ApJ...743L..16O, 2016A&A...595A..83E}.
Additional elevation is inferred in a number of disks by the strong lines of small carbon chains like \ce{C_2H} and \ce{C_3H_2} that require C/O ratios of 1.5-2 to form efficiently \citep{2016ApJ...831..101B,2019A&A...631A..69M,2019ApJ...876...25B,2021ApJS..257....9I,2021ApJS..257....7B}.
The elevated C/O ratio could be a result of the observed gas phase depletion of \ce{H_2O} and \ce{CO_2}  \citep{2010A&A...521L..33B,2011Sci...334..338H,2017ApJ...842...98D,2017A&A...601A..36B}.
Potentially direct photo-ablation of carbon rich grains can also contribute \citep{2021ApJ...910....3B}.
The C/O ratio in the planetary birth environment is one of the key ingredients in models of planetary atmospheres.
A slight increase in C/O ratio in a typical hot Jupiter atmosphere changes abundances of key carbon bearing species like \ce{CH_4} and \ce{C_2H_2} by multiple orders of magnitude \citep{2010fee..book..157L, 2019ARA&A..57..617M}.

\begin{table}[!b]
\vspace{-0.4cm}
\caption[]{LkCa~15 source properties}\label{tab:source_props}
\begin{tabular}{L{.35\linewidth}L{.28\linewidth}L{.23\linewidth}}
\hline\hline \noalign {\smallskip}
Parameter & Symbol & Value\\
\hline \noalign {\smallskip}
Right Ascension & RA (J2000) & 04:39:17.8\\
Declination & DEC (J2000) & +22:21:03.21\\
Distance & $d$ (pc) & 157.2\\
Spectral type & SpT & K5\\
Age & $t$ (Myr)	& 2 (0.9-4.3)\\
Effective Temperature & $T_{\rm eff}$ (K) &	4730\\
Stellar Luminosity & $L_{\star}$ (L$_{\odot}$)&	1.04 \\
Stellar Radius&$R_{\star}$ (R$_{\odot}$)& 1.0\\
Stellar Mass & $M_{\star}$ (M$_{\odot}$)&	1.03\\
Mass accretion rate & log$_{10}\dot{M}$ (M$_{\odot}$ yr$^{-1}$) & -8.7 \\
Visible Extinction & $A_{\rm V}$ (mag) & 1.5\\
Inclination & $i$ (deg)	& 55 \\
Position Angle & $PA$ (deg) & 60 	\\
Systemic velocity & v$_{\rm sys}$ (km s$^{-1}$) &	6.1 \\
\hline \noalign {\smallskip}
\end{tabular}
\textbf{References.} \citet{2010ApJ...717..441E,2015A&A...579A.106V,2020ApJ...890..142P,2021A&A...649A...1G}
\end{table}

Nitrogen bearing molecules are proposed to be good candidate gas mass tracers, since the main molecular carrier of N, \ce{N_2}, has such a low freeze-out temperature that depletion of nitrogen is likely less severe than that of oxygen and carbon \citep{2004A&A...414L..53T,2018ApJ...865..155C, 2018A&A...615A..75V}.
A smaller fraction of the total amount of nitrogen is contained in meteorites and comets, which suggests that less material is depleted from the gas phase by freezeout and further processing on the grains \citep[see more details in][]{2015PNAS..112.8965B,2019ApJ...881..127A}.
Since \ce{N_2} cannot be observed directly under typical disk conditions by lack of a dipole moment, much less abundant species like HCN, CN and \ce{N_2H+} are necessary to probe the disk mass.
However, most of these molecular abundances are dependent on the C/H abundance in the disk.
\ce{N_2H+}, which  is observed in a large number of disks \citep{2010ApJ...720..480O,2011ApJ...734...98O, 2019ApJ...882..160Q}, is the most promising candidate as it does not contain a carbon atom, but is destroyed by proton transfer in the vicinity of gaseous CO \citep{2017A&A...599A.101V,2019ApJ...881..127A}.
\citet{2019ApJ...881..127A, 2022ApJ...927..229A} show, based on a small survey of \ce{N_2H+} emission lines, that \ce{N_2H+} can be used in combination with \ce{C^18O} to constrain the carbon abundance and total gas mass, which is confirmed by the recent modeling work of \citet{2022arXiv220109900T} comparing the outcome with established HD modeling.
No single molecule is the optimum tracer, but we investigate if perhaps a combination of these molecules would work best to disentangle the gas mass from elemental abundance determination.

This paper is organized as follows.
In Sect. \ref{sec:observations} and \ref{sec:observational_results} the object of study, LkCa~15, previous work on measuring the gas mass in this system, and new observations are presented.
Sect. \ref{sec:data_analysis}  summarizes the physical-chemical modeling framework. 
Sect. \ref{sec:results} compares the results of a range of possible modeling parameters with the data, and in Sect. \ref{sec:discussion} we discuss the results and draw our conclusions.

\section{Observations}
\label{sec:observations}
\subsection{The LkCa~15 disk}
\label{sec:source}
In this paper we focus on one specific source, LkCa~15, that is one of the few disks for which observations in many gas mass tracers exist, including an HD upper limit.
Comparing multiple proposed gas mass tracers and the C/O ratio tracer \ce{C_2H} we aim to constrain the gas mass and C/H and O/H abundances in the bulk of the disk.
LkCa~15 is a relatively young \citep[$\sim$2 Myr old][]{2018ApJ...865..157A} T Tauri star located at a distance of 157.2 pc as determined by Gaia \citep{2016A&A...595A...1G,2021A&A...649A...1G} in the Taurus star forming region.
It has a well studied transition disk, with a dust disk cavity of $\sim$50-65 AU where the millimeter sized dust grains are depleted \citep[e.g.,][]{2006A&A...460L..43P,2012ApJ...747..136I,2020A&A...639A.121F}.
Most of the millimeter dust continuum is emitted from 2 bright rings between 50-150 AU.
Micron-sized dust grains are observed in scattered light observations in an inner disk inside the dust gap up to $\sim$30 AU \citep{2016PASJ...68L...3O,2016ApJ...828L..17T}.
Spatial segregation in dust sizes supports the presence of a massive planet orbiting at $\sim$40 AU \citep{2012A&A...545A..81P,2020A&A...639A.121F}.
There has been speculation about the existence of multiple planets inside the cavity based on asymmetric emission in scattered light and \ce{H\alpha} emission \citep{2012ApJ...745....5K,2015Natur.527..342S}, but these were later shown to be likely due to disk features \citep{2016ApJ...828L..17T,2019ApJ...877L...3C}.
The gas disk extends up to $\sim$1000 AU \citep{2019ApJ...881..108J}, and has a significantly smaller cavity than that in the dust of 15-40 AU \citep{2015A&A...579A.106V,2019ApJ...881..108J,leemker2021}.

Previous estimates of the mass in the LkCa~15 disk were based on the SED and the millimeter dust continuum \citep[$\sim$0.06 $M_\odot$][]{2012ApJ...747..136I,2015A&A...579A.106V,2020A&A...639A.121F} and combining the SED with the upper limit from the HD 1-0 line, which gives 0.062 $M_\odot$ \citep{2016ApJ...831..167M}.
\citet{2019ApJ...881..108J} used optically thick \ce{CO} and \ce{^{13}CO} lines and the assumption of a gas-to-dust ratio of 1000 and a constant CO/\ce{H_2} abundance of 1.4$\times10^{-4}$ to find a gas mass of $\sim$0.1$ M_\odot$. 
\citet{2020A&A...639A.121F} finds an upper limit of the gas-to-dust ratio using stability arguments against gravitational collapse of 62 and 68 in the two bright dust continuum rings, respectively, containing the bulk of the total dust mass.

\begin{table*}[!t]
\vspace{-0.2 cm}
\caption[]{LkCa~15 line properties}\label{tab:line_props}
\vspace{-0.2 cm}
\begin{tabular}{L{.11\linewidth}C{.09\linewidth}C{.13\linewidth}C{.1\linewidth}C{.09\linewidth}C{.075\linewidth}C{.07\linewidth}C{.06\linewidth}C{.07\linewidth}}
\hline\hline \noalign {\smallskip}
Line & Telescope & Program ID& Integrated flux & Beam size  & Beam PA & $\nu$ & $E_{\rm u}$& log$_{10}$$A_{\rm ul}$ \\
&&&(Jy km s$^{-1}$)&($^{\prime\prime}$)&(deg) &(GHz) & (K) &(s$^{-1}$)\\
\hline \noalign {\smallskip}
$^{12}$CO $J$=2-1&ALMA& 2018.1.01255.S &16 $\pm$ 2&0.34 x 0.25 & -9.7 & 230.538&	16.6&-6.16050\\
$^{13}$CO $J$=2-1&NOEMA&S20AT&5.0 $\pm$ 0.5&1.13 x 0.73 & 20.4 & \multirow{2}{*}{220.399}&\multirow{2}{*}{15.9}&	\multirow{2}{*}{-6.51752}	\\
$^{13}$CO $J$=2-1&ALMA&2018.1.00945.S&5.5 $\pm$ 0.5&0.35 x 0.25 & 26.7\\
$^{13}$CO $J$=6-5&ALMA&2017.1.00727.S&8.3 $\pm$ 0.9&0.32 x 0.30 & -31.2&661.067&111.1&	-5.02695	\\
C$^{18}$O $J$=2-1&NOEMA&S20AT&1.1 $\pm$ 0.1&1.14 x 0.74 & 20.4&\multirow{2}{*}{219.560}&\multirow{2}{*}{15.8}&\multirow{2}{*}{-6.22103}\\
C$^{18}$O $J$=2-1&ALMA&2017.1.00727.S&1.0 $\pm$ 0.1&0.36 x 0.27 & 26.6\\
C$^{17}$O $J$=2-1&NOEMA&S20AT&0.36 $\pm$ 0.07&1.13 x 0.72 & 20.9&224.714&16.2&	-6.19211\\
$^{13}$C$^{18}$O $J$=2-1&NOEMA&S20AT&< 0.17&1.17 x 0.78 & 21.5&209.419&15.1&-6.28123	\\
CN $N$=2-1,&\multirow{2}{*}{NOEMA}&\multirow{2}{*}{S20AT}&\multirow{2}{*}{8.1 $\pm$ 0.8}&\multirow{2}{*}{1.12 x 0.72}&\multirow{2}{*}{20.9}&\multirow{2}{*}{226.876}&\multirow{2}{*}{16.3}&\multirow{2}{*}{-3.94188}\\
\hspace{7mm}$J$=$\frac{5}{2}$-$\frac{3}{2}$\Bstrut\\
\ce{C_2H} $N$=3-2,&\multirow{2}{*}{ALMA}&\multirow{2}{*}{2016.1.00627.S}&\multirow{2}{*}{1.7 $\pm$ 0.2}&\multirow{2}{*}{0.59 x 0.49}&\multirow{2}{*}{-26.3}&\multirow{2}{*}{262.1}&\multirow{2}{*}{25.2}&	\multirow{2}{*}{-4.31202}\\ 
\hspace{7mm}$J$=$\frac{5}{2}$-$\frac{3}{2}$\Bstrut\\
\ce{C_2H} $N$=3-2,&\multirow{2}{*}{ALMA}&\multirow{2}{*}{2016.1.00627.S}&\multirow{2}{*}{2.4 $\pm$ 0.3}&\multirow{2}{*}{0.59 x 0.49}&\multirow{2}{*}{-26.3}&\multirow{2}{*}{262.0}&\multirow{2}{*}{25.2}&	\multirow{2}{*}{-4.27528}\\
\hspace{7mm}$J$=$\frac{7}{2}$-$\frac{5}{2}$\Bstrut\\
\ce{N_2H+} $J$=3-2 & ALMA&2015.1.00678.S& 1.82 $\pm$ 0.02 & 0.32 x 0.29& -23.0 & 279.511 & 26.8 & -2.86895\\
HD $J$=1-0&\textit{Herschel}&-& < 19.1&6.6&-&2674.99&128.4&	-7.26648\\
\hline \noalign {\smallskip}
\end{tabular}
\textbf{Notes:} Flux errors include 10\% calibration error.
\end{table*}

\subsection{Observational details}
\label{ssec:observations_details}
%Using rare CO isotopologue NOEMA data, high resolution ALMA CO data, the C/O tracers \ce{C_2H}, and nitrogen baring species CN and \ce{N_2H+} and the HD 1-0 upper limit we aim to constrain the gas mass and C and O elemental volatile abundances in the disk of LkCa~15.
We present new NOEMA (NOrthern Extended Millimeter Array) observations targeting the optically thin \ce{^13C^18O} and \ce{C^17O}~$J$=2-1 lines, \ce{^13CO}, \ce{C^18O} and CN. 
These new NOEMA observations combined with a large number of high-resolution archival ALMA CO isotopologue observations, \ce{N_2H+} and \ce{C_2H} emission, and a HD upper limit allow us to study the possible gas mass range and composition of LkCa15 in much more detail, breaking the degeneracy between the disk gas mass and C/H and O/H abundances.

The new NOEMA observations are taken on the 26th and 27th of November 2020. 
All nine antennae are used with baselines ranging from 32 to 344 m.
Three calibrators are used, 0507+179, LKHA101, and 3C84 to correct, respectively, the phase, amplitude, flux and radio interference.
We integrated for 7.5 hours on source resulting in a sensitivity of 40 mJy/Beam at a velocity resolution of 0.2~km~s$^{-1}$.
Imaging is done using \texttt{CASA} v5.4.0. \citep{2007ASPC..376..127M} using Briggs weighting and a robust factor of 0.5.
The resolution of the final images is typically 1.1$\dprime$~x~0.7$\dprime$ or $\sim150$ AU at the distance of LkCa~15.
Observational details are given in Table \ref{tab:line_props}.

Additionally, this project makes use of data from various archival ALMA programs, summarized in Table \ref{tab:line_props}.
We included high resolution CO~$J$=2-1, \ce{^{13}CO}~$J$=2-1 and \ce{C^{18}O}~$J$=2-1 lines to constrain the radial profile of the CO emission, \ce{^{13}CO}~$J$=6-5 to further constrain the temperature profile \citep{leemker2021}, and \ce{C_2H} to constrain the C/O ratio in the disk.
CO isotopologue ALMA data were self-calibrated in CASA and CLEANed using a Keplerian mask. 
Further details of the data and initial calibration are described in \citet{leemker2021}.
\ce{C_2H} ALMA observations and reduction are described in more detail in \citet{2019ApJ...876...25B}.
\ce{N2H+} observations are described in \citet{2019ApJ...882..160Q}, and are consistent with those reported in \citet{2020ApJ...893..101L}.
The upper limit for HD $J$=1-0 is taken at 3$\sigma$ from \textit{Herschel} data \citep{2016ApJ...831..167M}.
The properties of all emission lines and an overview of the observations are presented in Table \ref{tab:line_props}.

\subsection{Data analysis}
\label{ssec:data_analysis}
\begin{figure*}[!t]
    \centering
    \includegraphics[width = \linewidth]{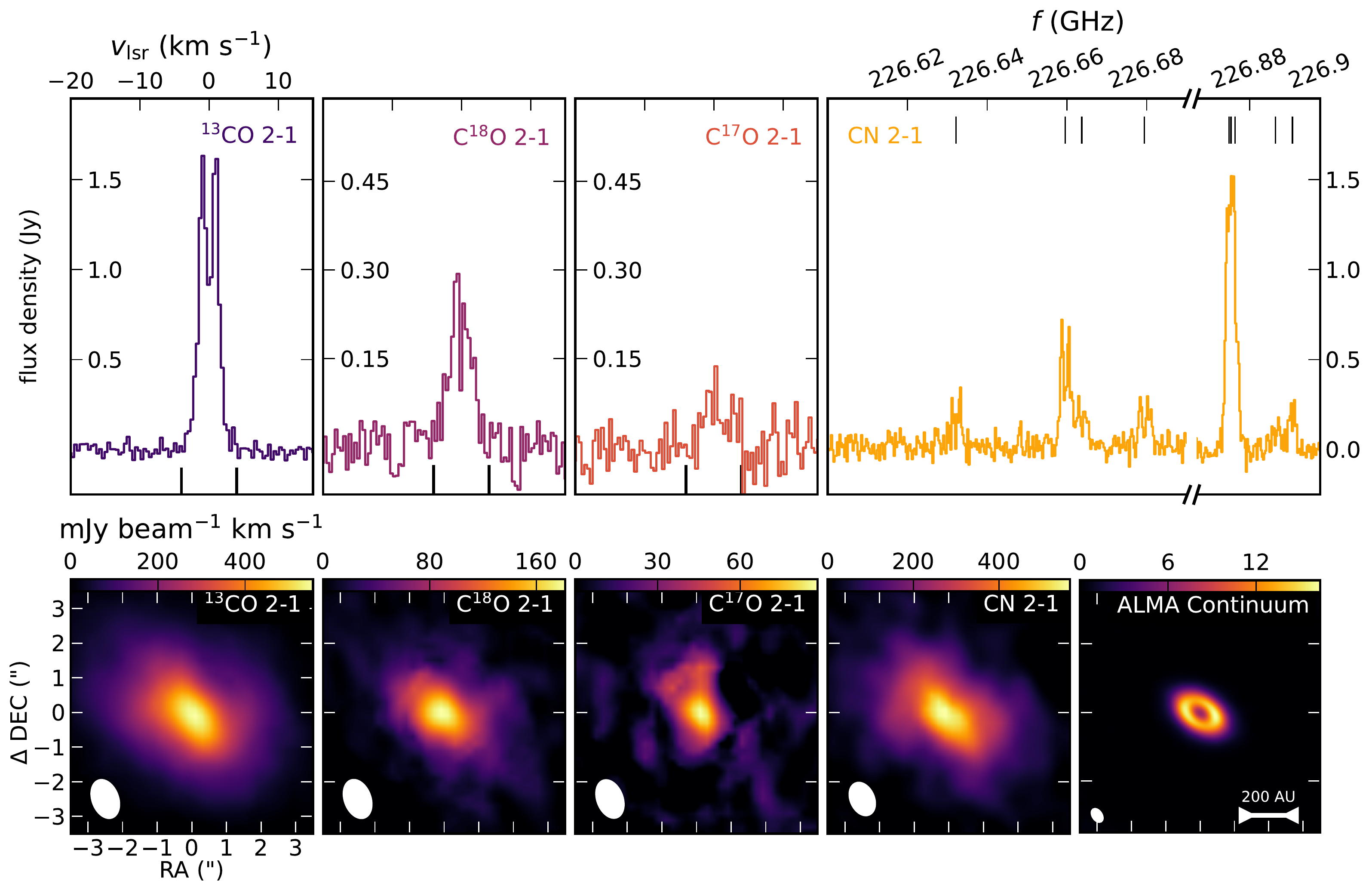}
    \caption{Disk integrated spectra (top) and total intensity or moment 0 maps of the new emission lines (bottom) obtained with NOEMA at 1.1$\dprime$x0.7$\dprime$. ALMA band 6 continuum data at 0.3$\dprime$x0.3$\dprime$ are shown for comparison. Integration limits for the moment 0 maps are indicated with black tickmarks in the bottom at -4 and +4 km s$^{-1}$. Robustly detected CN hyper-fine structure lines are indicated in the top right panel with black tick marks. For each moment 0 map the beam is shown as a white ellipse in the lower left corner.}
    \label{fig:mom0s}
\end{figure*}

\begin{figure*}[!t]
    \centering
    \includegraphics[width = \linewidth]{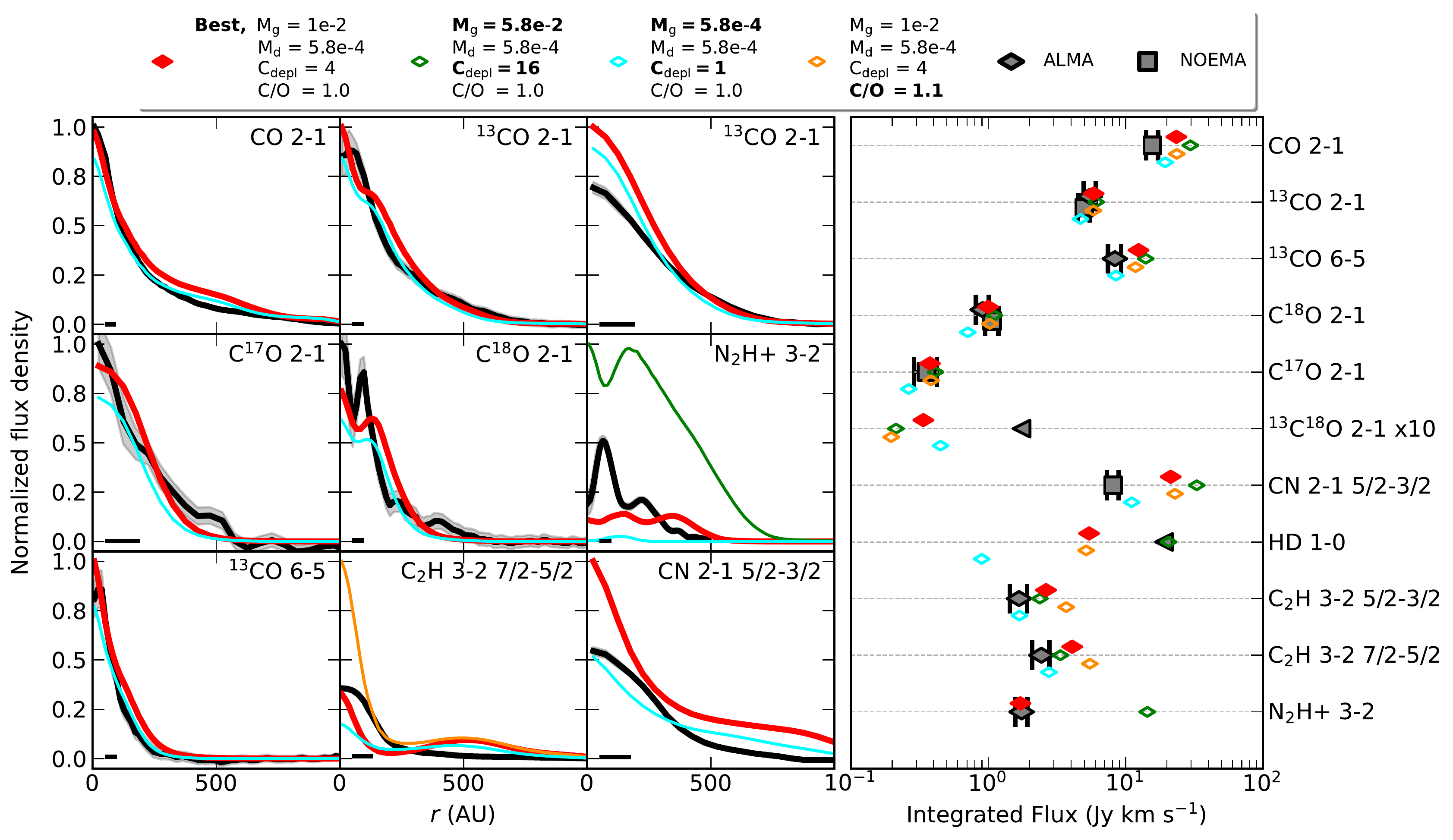}
    \caption{Left: Azimuthally averaged radial profiles of the different gas tracers used in this work. The data are shown as black lines with shaded 1$\sigma$ uncertainty, and normalized to the brightest point for visualization purposes. The full width at half maximum of the beams is shown as a black horizontal line in the lower left corner. The best fitting model with a gas mass of 0.012 $M_\odot$ and a factor four in carbon depletion is overplotted in red. The blue line shows the same model but with a lower gas-to-dust ratio, resulting in a gas mass of $4\times10^{-4}$M$_\odot$, and no carbon depletion. This model reproduces the optically thick CN and CO lines somewhat better, but underproduces the optically thin \ce{N_2H+} line. The orange line illustrates the \ce{C_2H} radial profile for an increased C/O ratio of 1.1. With C/O > 1 excess carbon, not in CO, changes the carbon chemistry completely.\\
    Right: integrated line flux of the various emission lines analyzed in this paper. Black squares represent the observed line fluxes of NOEMA data, black diamonds represent the observed line fluxes of ALMA data, black triangles denote upper limits, taken at 3$\sigma$. Colored diamonds show the results for the same models as in the left panel.}
    \label{fig:gas_radprofs}
\end{figure*}

Integrated emission or moment 0 maps were created for all NOEMA lines and CO isotopologue ALMA data by integrating over the emission inside the same Keplerian mask that is used in the CLEANing process, set by the source parameters given in Table \ref{tab:source_props} and restricted to $\pm$4 km s$^{-1}$ from the source velocity.
The moment 0 maps are presented in Fig.~\ref{fig:mom0s}, together with the \ce{^12CO} $J$=2-1 and continuum images of the ALMA Band 6 data for comparison.
Integrated line fluxes are determined using a mask extending to 6.5$\dprime$ or $\sim1000$ AU in radius and are presented in Table \ref{tab:line_props} and are shown in black in the right panel of Fig.~\ref{fig:gas_radprofs}.


All moment 0 maps are deprojected and averaged along azimuth using the geometrical source properties given in Table \ref{tab:source_props}.
We chose the source center for each molecule as the pixel (typically 1/5 of the beam) with the peak flux or, in case the cavity is resolved, the pixel with the lowest flux inside the cavity.
The source center is shifted with maximum 0.1$\dprime$ from the reference position, which is a result of the  observed eccentric inner disk \citep{2016ApJ...828L..17T,2016PASJ...68L...3O}, the high inclination (55$^\mathrm{o}$) of the main disk and the different observing dates.
The radial profiles are presented in Fig.~\ref{fig:gas_radprofs} as black lines, the 1 $\sigma$ uncertainty is masked around the curves.
We include the radial profile of the \ce{C_2H} emission from \citet{2019ApJ...876...25B} and the \ce{N2H+} from \citep{2019ApJ...882..160Q}. 
The radial profile of the continuum from \citet{2020A&A...639A.121F} is included in Fig.~\ref{fig:model_dust_setup}.

\section{Observational results}
\label{sec:observational_results}
All targeted CO isotopologues, except for \ce{^{13}C^{18}O}~$J$=2-1, are detected.
The spectra of all detected emission lines are presented in the top panel of Fig.~\ref{fig:mom0s}.
\ce{C^{17}O} $J$=2-1 is robustly detected with NOEMA at a $S/N$ of 6.
The \ce{^{13}C^{18}O} $J$=2-1 line is not detected so we listed the 3$\sigma$ upper limit in Table \ref{tab:line_props}.
The upper limit for \ce{^{13}C^{18}O} results in a minimum \ce{C^{17}O}/\ce{^{13}C^{18}O} ratio of 4.6, compared with the elemental abundance ratio of 24.
The observed \ce{^13CO} and \ce{C^18O}~$J$=2-1 disk integrated fluxes in the NOEMA and ALMA observations agree well with each other.
In addition we detect 8 hyper-fine CN~$N$=2-1 lines that we combine into one single image.
Based on the ratio of the hyper-fine lines we find that the emission is optically thick for the blended lines at 226.87 GHz and 226.66 GHz, but optically thin for the other lines.

Most of the observed CO isotopologue lines are centrally peaked, without resolving the inner cavity.
The inner cavity is resolved in the \ce{^{13}CO} $J$=2-1 and \ce{C^{18}O} $J$=2-1 ALMA data, lately confirmed by \citet{leemker2021} using \ce{^{13}CO} $J$=2-1 data with higher spatial resolution and using the high velocity line wings that trace regions closer in to the star than the spatial resolution of the data \citep{2021ApJS..257...15B}.
The CN emission is centrally peaked and extends out to 1000 AU, similar to the \ce{^{12}CO} data.

\section{Modeling framework}
\label{sec:model_analysis}
Our aim is to distentangle the gas mass and elemental abundances C/H and O/H in LkCa~15. 
For that purpose we made a representative model of LkCa~15 using the azimuthally symmetric physical-chemical model DALI \citep[Dust And LInes;][]{2012A&A...541A..91B,2013A&A...559A..46B}.
To find the best representative geometry, density structure and composition of the LkCa~15 disk we combined information from the Spectral Energy Distribution (SED), high resolution ALMA continuum observations, the emission lines discussed in Sect. \ref{ssec:observations_details} and observables from scattered light observations, near-infrared spectra from the Infrared Telescope Facility (IRTF) facility using SpeX \citep{2010ApJ...717..441E}.

\subsection{Model parameters}
\label{ssec:model_params}
\begin{figure*}
    \centering
    \includegraphics[width = \linewidth]{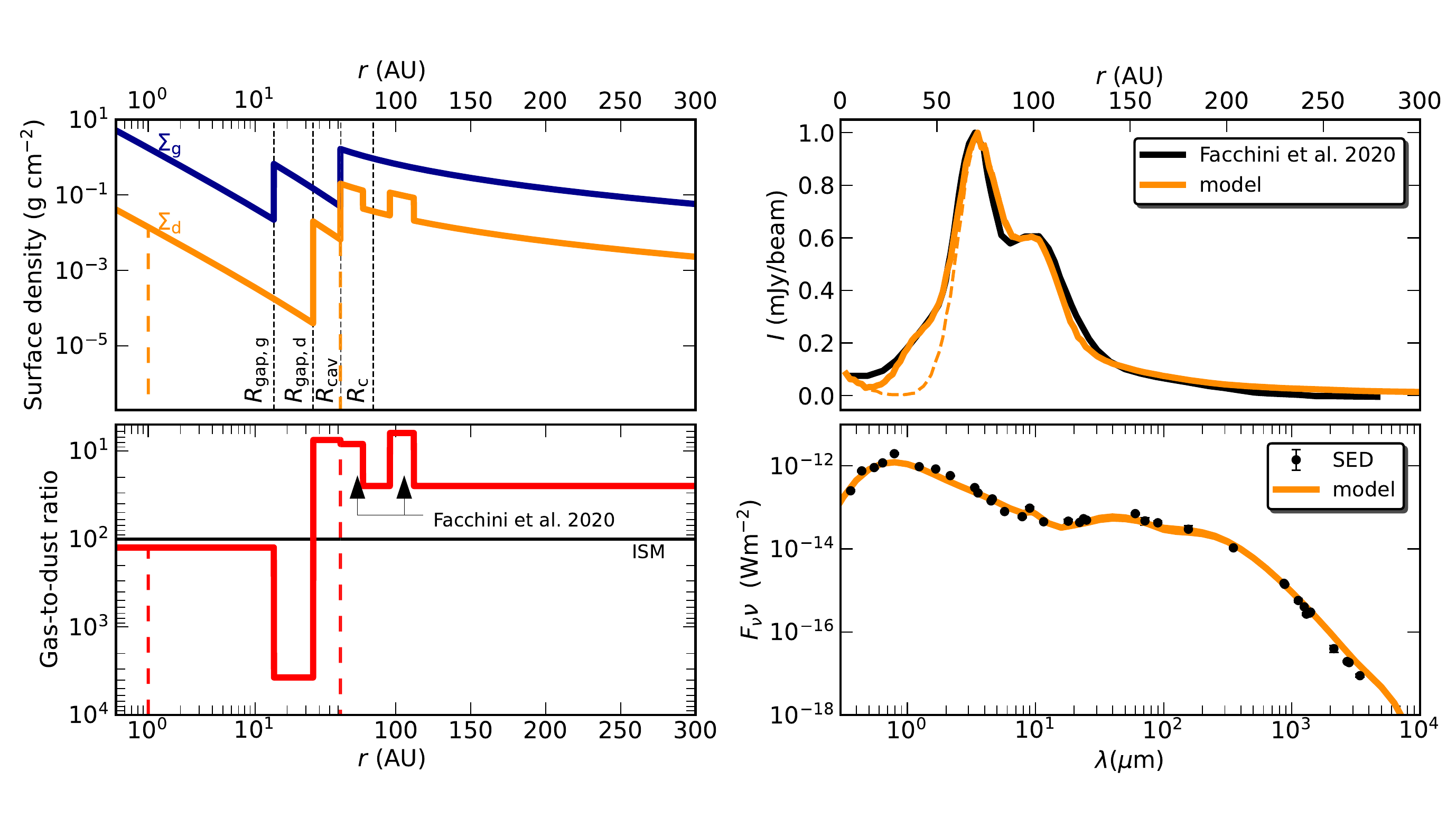}
    \caption{Setup of the DALI model for LkCa~15, Top left: normalized surface density of the dust (orange) and gas (blue) as function of radius, Bottom left: Gas-to-dust ratio as function of radius in the fiducial model, this is a direct consequence of the used gas and dust surface density profiles shown in the top left panel, Top right: radial profile of the continuum observations (black) and the model (orange) normalized to 1, Bottom right: observed SED in black with the modeled continuum fluxes in orange. Dashed lines show the setup for the same model with cleared cavity, discussed in App. \ref{app:paradependence}.}
    \label{fig:model_dust_setup}
\end{figure*}

The first step in our modeling was to set up a geometrical density structure including all radiation sources that is consistent with the data.
For the dust density structure we adopted a fully parametrized surface density profile with a power law surface density dependence on radius ($r$) and an exponential outer taper:
\begin{equation}
\label{eq:surface_density_profile}
\Sigma_{\mathrm{dust}}=\frac{\Sigma_{\mathrm{c}}}{\epsilon} \left(\frac{r}{R_{\mathrm{c}}}\right)^{-\gamma} \exp \left[-\left(\frac{r}{R_{\mathrm{c}}}\right)^{2-\gamma}\right],
\end{equation}
where $\Sigma_\mathrm{c}$ is the surface density at the characteristic radius, $R_\mathrm{c}$, $\gamma$ the power law index and $\epsilon$ the gas-to-dust ratio.
Following the approach of \citet{2006ApJ...638..314D} we split the dust into a small dust population which ranges from 0.005 to 1 micron and a large dust population which includes the small dust sizes, but extends to 1000 micron.
To calculate the opacities we assume a standard ISM dust composition following \citet{2001ApJ...548..296W}, combined using Mie theory, consistent with \citet{2011ApJ...732...42A}.
We also included polycyclic aromatic hydrocarbons (PAHs), assumed to be 0.1 \% of the ISM abundance following \citet{2006A&A...459..545G}.
The small dust grains follow an exponential dependence on the height in the disk, given by
\begin{equation}
    \label{eq:small_grains_distribution}
    \rho_{\mathrm{d}, \mathrm{small}}=\frac{(1-f_\mathrm{\ell}) \Sigma_{\mathrm{dust}}}{\sqrt{2 \pi} r h} \exp \left[-\frac{1}{2}\left(\frac{\pi / 2-\theta}{h}\right)^{2}\right],
\end{equation}
where $\theta$ is the opening angle from the midplane as seen from the center, $h$ is the scale height defined by $h=h_{\mathrm{c}}\left(r/R_{\mathrm{c}}\right)^{\psi}$ and $f_\mathrm{\ell}$ is the mass fraction of large grains. 
We parametrized dust settling by replacing the scale height of the large grains by $\chi h$, so that only a fraction 1-$f_\mathrm{\ell}$ of the grains is distributed throughout the height in the disk.
We fixed $\chi$ at 0.2 and $f_{\mathrm{\ell}}$ at 0.98, adopted from \citet{2016ApJ...831..167M} \citep[see also][]{2011ApJ...732...42A}.
The gas surface density follows the same radial and vertical distribution as the small dust grain population, but scaled by the total gas-to-dust ratio, which we take initially at 100.
Within the dust sublimation radius, both dust and gas surface density are set to zero.
To include a cavity in the dust and gas we scaled the dust and gas surface density profiles independent from each other inside a given radius, $R_\mathrm{gap}$, with respect to the distribution given in Eq. \ref{eq:small_grains_distribution} with constant scaling factors $\delta_\mathrm{d}$ and $\delta_\mathrm{g}$.
To account for the three rings observed in the high-resolution continuum images \citep{2020A&A...639A.121F} we scaled up the dust density distribution in three intervals, see Fig.~\ref{fig:model_dust_setup} for a visualization of the density distribution of the dust and gas.

\subsubsection{Radiation field}
The stellar spectrum is approximated as a blackbody with temperature $T_\mathrm{eff}$~=~4730 K.
We added an additional UV component to account for gas accretion using a plasma temperature ($T_\mathrm{acc}$) of 10,000 K, with a luminosity given by 
\begin{equation}
    L_{\mathrm{acc}}(v)=\pi B_{v}\left(T_{\mathrm{acc}}, v\right) \frac{G M_{*}\dot{M}}{R_{*}} \frac{1}{\sigma T_{\mathrm{acc}}^{4}}
\end{equation}
where $B_{v}\left(T_{\mathrm{acc}}, v\right)$ is Planck's law for blackbody radiation, $M_{*}$ is the stellar mass, ${R_{*}}$ is the stellar radius and $\dot{M}$ the mass accretion rate.
We adopted stellar parameters from \citet{2010ApJ...717..441E} and \citet{2020ApJ...890..142P}, given in Table \ref{tab:source_props} resulting in L$_\mathrm{UV}$ = 0.06 L$_\odot$. 
Note that the observed values of $\dot{M}$ are uncertain up to an order of magnitude and can change over time.
We included X-rays assuming an X-ray luminosity of 10$^{30}$ erg s$^{-1}$ at a plasma temperature of 7 x $10^{7}$ K.
The cosmic ionization rate incident on the surface of the disk is set to 5 x $10^{-17} \mathrm{s}^{-1}$.

Using the sources of radiation and the density structure described in Sect. \ref{ssec:model_params}, we determined the dust temperature and continuum mean intensity in each grid cell using a Monte-Carlo approach, where photon packages are started randomly from the star, dust grains or the background.

\subsection{Chemical networks}
The gas temperature is iteratively solved together with the chemistry, since the gas temperature is dependent on the abundance and excitation of different coolants in the disk.
We used two different chemical networks in DALI to cover both the isotope selective processes of CO and proper treatment of \ce{C_2H} and CN.
For the CO isotopologue emission we used the HD chemistry described in \citet{2017A&A...605A..69T}, but adapted it to the reduced CO isotopologue network described in \citet{2016A&A...594A..85M} rather than the full network described in \citet{2014A&A...572A..96M} to save computation time.
This network includes HD, D, \ce{HD+}, \ce{D+}, \ce{^{13}CO}, \ce{C^{18}O}, \ce{C^{17}O}, and \ce{^{13}C^{18}O} as separate species, resulting in a total of 189 species and 5789 reactions  \citep[compared to 280 species and 9789 reactions in][]{2017A&A...605A..69T}.
For CN and \ce{C_2H} we used the expanded network described in \citet{2018A&A...609A..93C} and \citet{2018A&A...615A..75V}. Nitrogen isotopes are not considered.
Reaction types include standard gas-phase reactions, photoionization and -dissociation, X-ray and cosmic ray induced reactions, freeze-out and (non-) thermal desorption, PAH and small grain charge exchange, hydrogenation and reactions with vibrationally excited \ce{H_2} (H$_{2}^{*}$).
Details of these reactions are described in \citet{2012A&A...541A..91B}, \citet{2017A&A...605A..69T} and \citet{2018A&A...615A..75V}.

We adopted fiducial ISM gas-phase volatile elemental carbon, oxygen and nitrogen abundances of C/H~=~135 ppm, O/H~=~288 ppm and N/H~=~21.4 ppm, respectively \citep{1996ApJ...467..334C,1998ApJ...493..222M,2012ApJ...760...36P}, and $^{13}$C, $^{18}$O, $^{17}$O abundance ratios of 77, 560, and 1792 with respect to their most abundant isotopes \citep{1996ApJ...467..334C,1998ApJ...493..222M,2012ApJ...760...36P,1999RPPh...62..143W}.
All volatile carbon, oxygen and nitrogen starts in atomic form in the gas, but can cycle between gas and ice.
The chemistry is run time dependently for 1 Myr. Most line strengths do not change when run in steady-state mode, because the surface layer dominates the abundance of these molecules where the chemistry is fast and emission is not sensitive to the time step.

\ce{N_2H+} is modeled using a separate, simple chemical network described in \citet{2017A&A...599A.101V} for a better treatment of the charge distribution.
This network focuses specifically on the ionization balance between \ce{H_3+}, \ce{N_2H+} and \ce{HCO+}, including freezeout, thermal desorption and photodissociation of CO and \ce{N_2}. 
Using this simplified network has the advantage of being easier to understand the ionization balance, while using the same temperature structure and CO and \ce{N_2} abundances as in the complete model.

After the chemistry is converged, the model is ray-traced in the analyzed transition lines and at 200 continuum wavelengths from 0.1-10,000 micron. Following a similar approach as described in Sect. \ref{ssec:observations_details} we create moment 0 maps and radial profiles for each of the model lines.

\subsection{Modeling approach}
Initially we fit the disk parameters by eye in a logical order, that we describe in the following sections, so that they converge to a model that best represents the available data.
The large number of parameters and the computational time of the models make it impossible to use $\chi^2$ fitting or Monte Carlo processes on the whole parameter space.
This means that it is not possible to derive formal uncertainties of model parameters and to determine correlations or degeneracies between parameters.
The final best fit parameters are summarized in Table \ref{tab:model_props}.

\begin{figure}[ht]
    \centering
    \includegraphics[width = 0.5\textwidth]{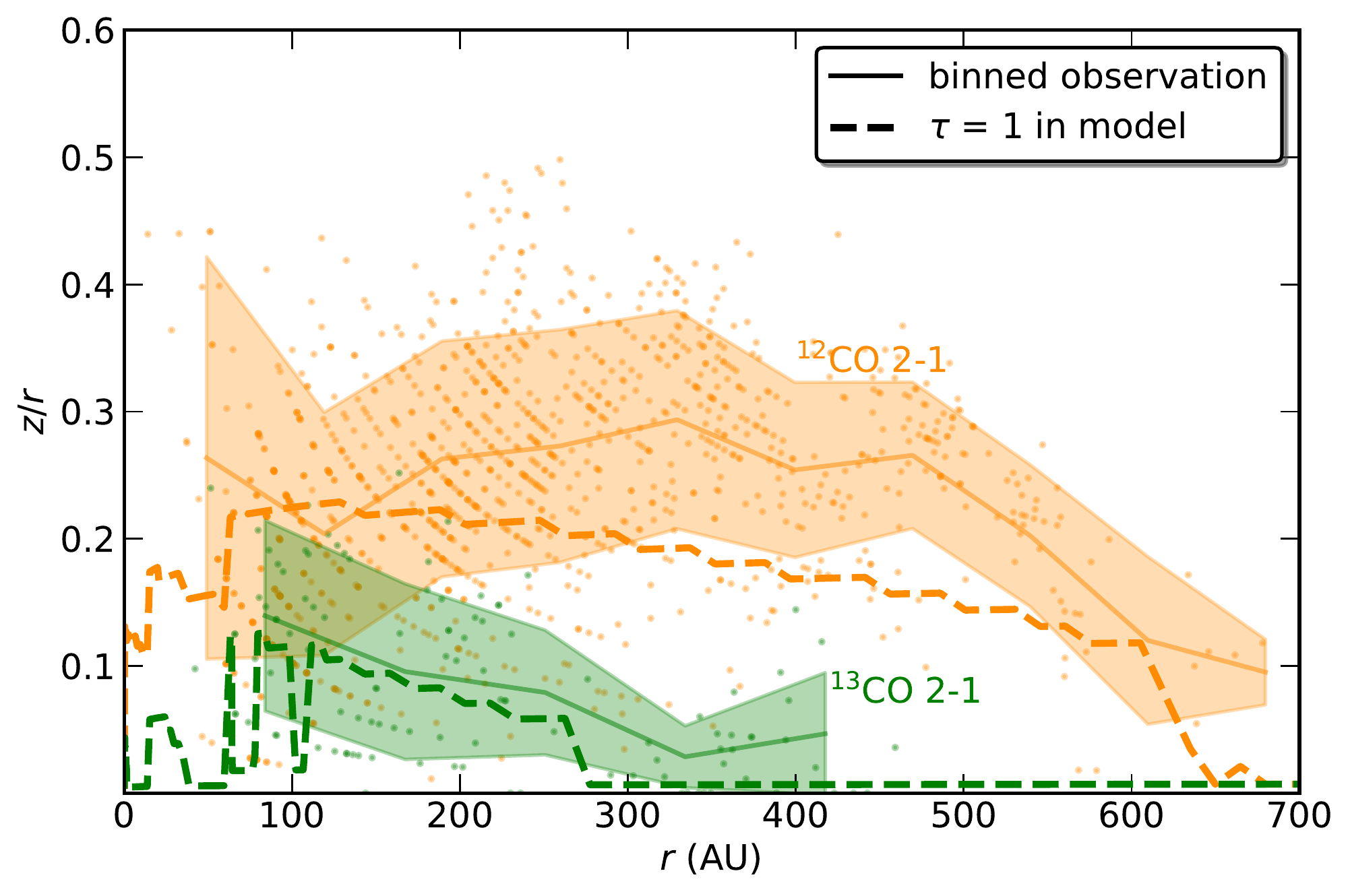}
    \caption{Emitting heights of the \ce{$^{12}$CO} (orange) and \ce{^13CO} (blue) $J$=2-1 emission, taken from \citet{leemker2021}. Scattered points represent the emitting height for each channel, whereas the solid line represents a radially binned average with 1$\sigma$ error. The dashed line shows the height of the $\tau=1$ line in the DALI model. The disk is flat in both the observations as well as the model.}
    \label{fig:emission_height}
\end{figure}

\subsubsection{Fitting the dust density distribution}
The first step was to find a dust density distribution that agrees well with the data.
For that purpose we compared the modeled continuum spectrum with the observed SED and the radial profile of the modeled emission to the radial profile of the high-resolution 1.35 mm continuum data \citep{2020A&A...639A.121F}.
The SED is taken from \citet{2013ApJ...771..129A} and dereddened using the CCM89 extinction curve \citep{1989ApJ...345..245C} and the observed visible extinction $A_\mathrm{V}$=1.5 \citep{2010ApJ...717..441E}.
As a starting point we used the model by \citet{2015A&A...579A.106V}, but updated to the new GAIA DR 3 distances and most recent observations.
We first fit the aspect ratio, $h_\mathrm{c}$, by matching the model with the observations at the base height of the silicate feature at 10 micron, assuming grain size distributions given in Sect. \ref{ssec:model_params}.
This aspect ratio agrees well with the observed emitting heights of the optically thick \ce{CO} and \ce{^13CO} lines, see Fig.~\ref{fig:emission_height}.
Subsequently we fit the dust gap radius based on the radial profile of the continuum observations, using an inner disk that extends to 35 AU as revealed by scattered light observations \citep{2016PASJ...68L...3O,2016ApJ...828L..17T}.
We then varied the flaring index, $\psi$, and the dust depletion in the gap, $\delta_\mathrm{d}$, to roughly fit the SED and continuum radial profile.
The flaring index is low, $\psi$ = 0.075, consistent with a flat disk.
The dust mass is fixed to $5.8\times10^{-4}$ M$_\odot$ based on the lower limit from the continuum observations \citep{2012ApJ...747..136I}. This dust mass agrees well with the slope of the SED beyond 1 mm (see Fig.~\ref{fig:model_dust_setup}), given the adopted grain size distributions and ISM-like grain composition.
We included three rings to account for the observed continuum rings, and varied the height of these to match the radial profile of the continuum observations.
The way these dust rings are treated results in a decreased model gas-to-dust ratio in these rings, consistent with the upper limits of 68 and 62 constrained by \citet{2020A&A...639A.121F}.
The cavity is not completely empty in our model, to account for the third continuum ring inside the dust cavity observed in the 1.3 mm continuum.
Clearing a large cavity from dust between 35 and 63 AU (dashed lines in Fig.~\ref{fig:model_dust_setup}) does not significantly change the gas temperature structure, and has no significant effect on the line emission as we show in Appendix \ref{app:paradependence} \citep[see also the discussion in][]{2015A&A...579A.106V}.
The parameters of the most representative model are listed in Table \ref{tab:model_props}.

\subsubsection{Fitting the gas density distribution}
\label{sssec:gasdistribution_fit}
The second step in constraining the model is determining the gas density distribution, with
$\gamma$ and $R_\mathrm{c}$ constrained by the CO isotopologue lines.
For that purpose we ran a grid of models ranging between 50 < $R_\mathrm{c}$ < 150 and 0.5 < $\gamma$ < 4 and took the parameters that reproduced the radial profiles of the CO isotopologue lines the best beyond 200 AU.
The gas gap is constrained by high resolution \ce{^{13}CO} and \ce{C^{18}O} observations with ALMA.
The gas surface density inside the gap at 15 AU ($\sim0.1$ g cm$^{-1}$; see Fig.~\ref{fig:model_dust_setup}) is consistent with upper limits on the total column density determined using \ce{C^18O} \citep{leemker2021}, but is an order of magnitude higher than the column density upper limit inside 0.3 AU determined from ro-vibrational CO upper limits \citep{2009ApJ...699..330S}. 
Lowering the gas surface density in the gap further results in underproducing \ce{C^17O} and \ce{^12CO} inside the gas gap.
We find that LkCa~15 is a large, flat, massive disk with significant structure inside 100~AU in both gas and dust.

We used the \ce{C_2H} flux to determine the C/O ratio in the disk.
The elemental carbon depletion and gas mass are then based on the CO isotopologue data, \ce{N_2H+}, CN and the HD upper limit.
To study the degeneracy between the carbon abundance and the gas mass, we run a grid with models for gas-to-dust ratios of [205, 100, 63, 39, 25, 15, 10, 4, 2, 1] and carbon depletion factors [1, 2, 4, 8, 16, 32].
This results in a range of gas masses between 0.1~M$_\odot$ and 5$\times10^{-4}$~M$_\odot$. We keep the dust mass fixed at a value of 5.8$\times10^{-4}$~M$_\odot$.
We allowed a modeled emission line uncertainty by a factor of two to find a confined region where the model is consistent with the data.
Confining the region to the model uncertainties circumvents the dependence of the best fit on the $S/N$ of the lines, since the brightest lines (e.g., CO and CN) are not the most sensitive to variations in gas mass and carbon abundance.
A factor of two is motivated by earlier work from \citet{2016yCat..35880108K} where they vary the most impacting modeling parameters to determine the effect on the total line flux for CO and isotopologues. 
Also, \citet{2012A&A...541A..91B} and \citet{2013A&A...559A..46B} in their benchmark of the DALI model find deviations in flux within a factor of two compared to similar modeling codes due to uncertainties in the reaction rates and desorption temperatures.
The geometrical structure of the disk is constrained in this work due to the wealth of data available for LkCa~15.
More specific uncertainties and the dependence of the molecular emission flux on important parameters are described in Sect. \ref{sec:discussion}.
In App. \ref{app:paradependence} we present the result of variations in key parameters and their influence on the total line flux.

\begin{table}[ht]
\vspace{-0.2 cm}
\caption[]{LkCa~15 best fit model parameters}\label{tab:model_props}
\vspace{-0.2 cm}
\begin{tabular}{L{.46\linewidth}L{.19\linewidth}L{.21\linewidth}}
\hline\hline \noalign {\smallskip}
Parameter & Symbol & Value\\
\hline \noalign {\smallskip}
Dust sublimation radius         & $R_\mathrm{subl}$ (AU)                    & 0.08\\
Outer grid radius               & $R_\mathrm{out}$ (AU)                     & 1000\\
Characteristic scale height     & $h_\mathrm{c}$                            & 0.08\\
Flaring index                   & $\psi$                                    & 0.075\\
Characteristic radius           & $R_\mathrm{c}$ (AU)                       & 85\\
Density power law coefficient   & $\gamma$                                  & 1.4\\
Settling parameter              & $\chi$                                    & 0.2\\
Mass fraction large grains      & $f_\mathrm{\ell}$                            & 0.98\\
\textbf{Gas surface density}    & \textbf{$\Sigma_\mathrm{c}$ (g cm$^{-2}$) }  & \textbf{2.7}\\
\textbf{Total gas mass  }       & \textbf{$M_\mathrm{g}$ (M$_\odot$) }         & \textbf{1.2e-2}\\
\textbf{Gas-to-dust ratio }     & \textbf{gdr}                                 & \textbf{15}\\
Total dust mass                 & $M_\mathrm{d}$ (M$_\odot$)                & 5.8e-4\\
Dust gap radius                 & $R_\mathrm{gap,d}$ (AU)                   & 35\\
Dust gap depth                  & $\delta_\mathrm{d}$                       & 2e-4\\
Gas gap radius                  & $R_\mathrm{gap,g}$ (AU)                   & 15\\
Gas gap depth                   & $\delta_\mathrm{g}$                       & 1e-4\\
Cavity radius                   & $R_\mathrm{cav}$ (AU)                     & 63\\
Dust cavity depth               & $\delta_\mathrm{cav,d}$                   & 1e-1\\
Gas cavity depth                & $\delta_\mathrm{cav,g}$                   & 3e-2\\
Ring 1                          & $R_\mathrm{ring1}$ (AU)                   & 63-78\\
Scaling in ring 1               & $\delta_{r1}$                             & 3\\
Ring 2                          & $R_\mathrm{ring2}$ (AU)                   & 96-112\\
Scaling in ring 2               & $\delta_{r2}$                             & 4\\
\hline \noalign {\smallskip}
X-ray luminosity                & $L_\mathrm{X}$ (erg s$^{-1}$)             & 1e30\\ 
X-ray temperature               & $T_\mathrm{X}$ (K)                        & 7e7\\
Cosmic ray ionization rate      & $\zeta$ (s$^{-1}$)                        & 5e-17\\

\hline \noalign {\smallskip}
Carbon abundance                & C/H                                       & 3.4e-5\\
Oxygen abundance                & O/H                                       & 3.4e-5\\
C/O ratio                       & C/O                                       & 1\\
\hline \noalign {\smallskip}
\end{tabular}
\noindent\textbf{Notes.} bold text indicates the parameters that are varied in the 2D grid
\end{table}

\begin{figure*}[!t]
    \centering
    \includegraphics[width = \linewidth]{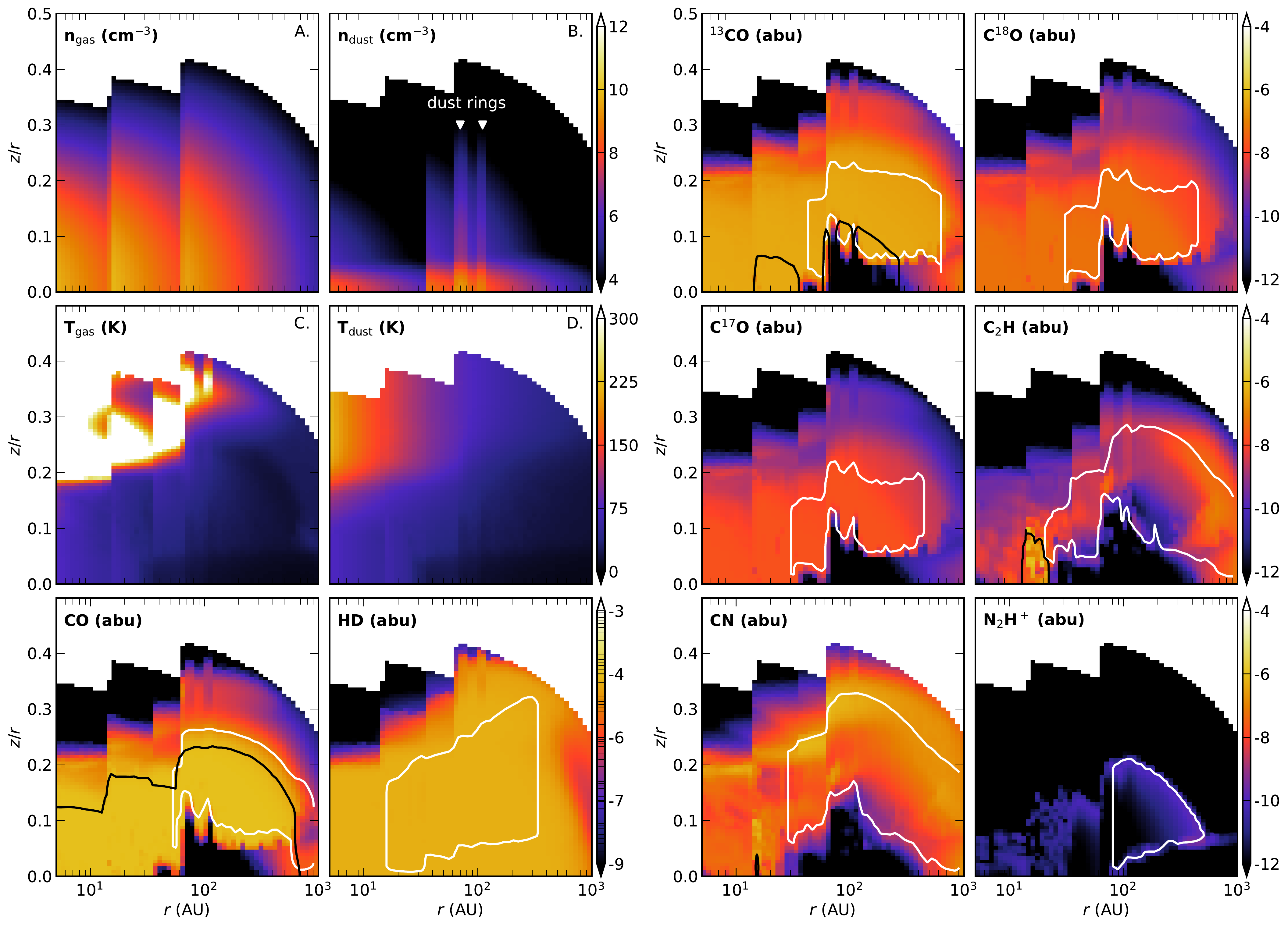}
    \caption{Overview of the DALI model output. The four panels left top (A-D) present the gas and dust density structure and the gas and dust temperature, respectively. The other panels present the molecular abundances of the molecular transitions in the most representative DALI model, assuming a gas-to-dust ratio of 25 and a carbon depletion factor of 4. The white contours represent the region where 95\% of the emission originates, the black lines show the $\tau~=~1$ surface for the optically thick emission lines.}
    \label{fig:dali_overview}
\end{figure*}

\begin{figure*}
    \centering
    \includegraphics[width = \linewidth]{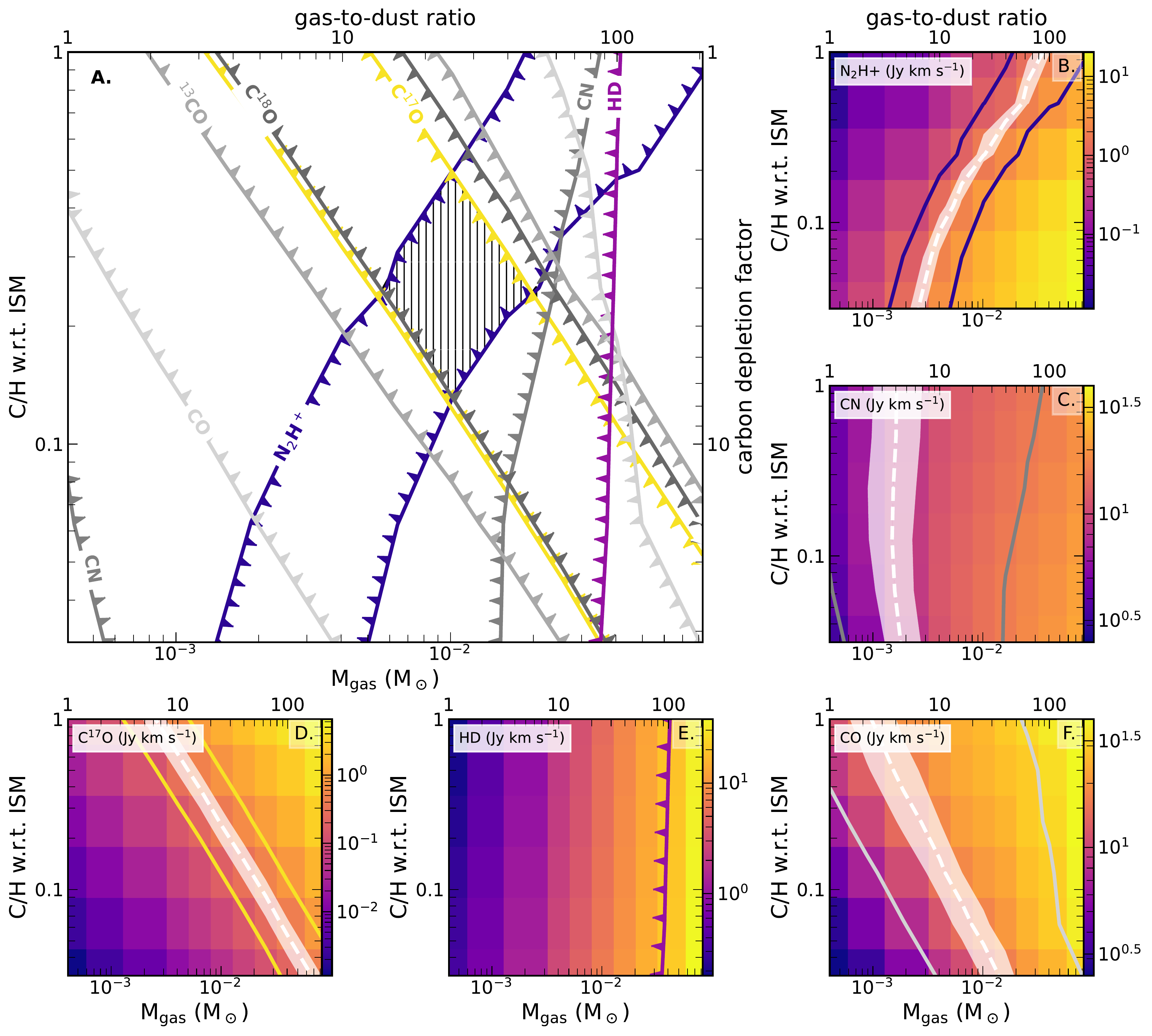}
    \caption{Results of the 10 x 6 model grid with carbon abundance versus disk gas mass. Panel A shows the limits on C/H and M$_{\rm gas}$ for each emission line, assuming a model uncertainty of a factor of 2. The hashed region denotes the parameter space that matches with all the observed emission lines. Panels B.-F. show the line flux of key C/H abundance and gas mass tracers \ce{N_2H+}, \ce{CN}, \ce{C^{17}O}, HD and \ce{^{12}CO}, respectively, for the same grid of models. The white contours depict the observed flux with observational uncertainty in shaded region. The colored contours mark an average deviation of the model by a factor of 2, the same as in the main panel. The HD observation denotes an upper limit.}
    \label{fig:gridplot}
\end{figure*}

\section{Results}
\label{sec:results}
\subsection{Gas distribution}
\label{ssec:results gasdistribution}
Fig.~\ref{fig:gas_radprofs} presents the radial profiles and the total fluxes of the best fitting model.
We find that a characteristic radius of $R_\mathrm{c}=$ 85~AU and $\gamma$~=~1.5 fit the emission best at large radii.

\subsubsection{HD}
The HD 1-0 emission in the model is confined to the inner region $<200$ AU (see Fig.~\ref{fig:dali_overview}), where the gas temperature is high enough \citep[30-70 K, see also][]{2017A&A...605A..69T} to populate the higher levels.
This relatively small emitting region is a direct result of the flat geometry of the disk, which means that regions further out in the disk are relatively cool.
This makes the total HD 1-0 flux susceptible to temperature variations inside the cavity at 63 AU and changes in the level of gas depletion with respect to the dust.

\subsubsection{CO isotopologues}
Using the high resolution CO isotopologue data we find that the gap inside 35 AU is less depleted in gas than in dust, but a significant gas depletion of a factor of $3\times10^{-2}$ throughout the cavity is necessary to explain the central dip observed in \ce{^{13}CO} and \ce{C^{18}O}.
The inner disk inside 15 AU is depleted further in gas up to a factor $10^{-3}$ (see Fig.~\ref{fig:model_dust_setup}).
This is consistent with earlier findings in \citet{leemker2021}; they report a gas cavity size $\sim$15 AU based on the line wings and radial profile of high $S/N$ CO isotopologue data.
The two steps in gas depletion (see the top left panel of Fig.~\ref{fig:model_dust_setup}) that were necessary to fit the CO lines could be a result of gradual gas depletion inside the cavity.

The best fitting model reproduces \ce{^{13}CO 2-1}, \ce{C^{18}O 2-1}, \ce{C^{17}O 2-1} and \ce{^{13}CO 6-5} within the 1$\sigma$ error at most radii.
The \ce{^13CO} and \ce{C^{18}O} emission is predicted to be moderately optically thick in the region just outside the gas cavity, but the \ce{C^{17}O} emission is optically thin throughout the disk.
This is consistent with the observations that show that the disk integrated \ce{C^18O}/\ce{C^17O} ratio of 3.1 is consistent with the expected ISM abundance ratio of 3.2.
The models are unable to reproduce the small bump at 500 AU observed in both the \ce{C^{17}O} and \ce{C^{18}O} radial profiles (see Fig.~\ref{fig:gas_radprofs}). 
This enhancement may be caused by non-thermal desorption processes of CO from the ice due to enhanced UV penetration or a slight change in C/H ratio as a result of infalling pristine material.

The \ce{^{12}CO} line is highly optically thick throughout the largest part of the disk (see Fig.~\ref{fig:dali_overview}) and thus primarily sensitive to temperature.
Our model reproduces the radial profile of this line and is consistent with the vast size of the disk (see Fig.~\ref{fig:gas_radprofs}).
The \ce{^{12}CO} model is somewhat too bright in the outer regions of the disk between 300-700 AU, by about a factor of 1.5.
The largest recovered angular scale of the \ce{^{12}}CO data is $r \sim600$ AU, which likely explains part of the flux that is missing in the outer regions.
Changing the geometry (i.e., $h_\mathrm{c}$ and $\psi$) does not change the temperature structure sufficiently.
The disk is flat (see Fig.~\ref{fig:emission_height}), and changing $h_\mathrm{c}$ will only marginally affect the temperature of the highest layers in the disk where the CO and CN emission originate.
Lowering the gas mass of the system lowers the height of the emitting layer and results in less flux contribution across the disk, matching the data better.
Altering the settling parameters $\chi$ or $f_\mathrm{\ell}$ (see Eq. \ref{eq:small_grains_distribution}), as suggested by \citet{2016ApJ...831..167M}, changes the temperature in the upper layers and would allow for better fitting of \ce{^12CO}. 
However, this results in decreased temperatures that would lead to unrealistically low CO isotopologue fluxes, even with increased values for $\gamma$ and ISM C/H abundance.

\subsubsection{N-bearing species}
\label{ssec:gasdistr_nbearing_species}
The \ce{N2H+} emission is confined to the inner 300 AU and is in general well reproduced by the model.
\ce{N2H+} is predominantly formed in regions where the \ce{N2} abundance dominates the CO abundance, for example as a result of differences in their desorption temperature or photo-dissociation rate.
\citet{2019ApJ...882..160Q} show that the emission can be separated in an extended component, attributed to the small layer between the (vertical) \ce{N2} and CO snow surfaces, and an additional ring at $\sim50$ AU due to the difference between the (radial) \ce{N2} and CO snowline.
The ring at $\sim50$ AU is underproduced in our model (see Fig.~\ref{fig:gas_radprofs}), which is likely because of the steep temperature gradient just outside the cavity.
The difference in desorption temperature between \ce{N2} and CO is only 3-7 K \citep{2006A&A...449.1297B} which means that the location of the CO and \ce{N_2} snowline cannot be determined accurate enough due to the uncertainty in the temperature structure and cavity properties of the system.
Using the location of the brightest ring in the \ce{N2H+} emission, \citet{2019ApJ...882..160Q} find that the midplane CO snowline is located at 58$^{+6}_{-10}$ AU.
This value is consistent with the location of the midplane CO snowline in our model. 
Assuming a CO freezeout temperature of 20 K we find that the snowline in our model is located at 66 AU, just outside the cavity.
Additional X-ray ionization close to the star, which is not taken into account in our modeling, and a locally decreased CO abundance closer to the snowline  \citep[see][]{2020ApJ...899..134K} could also partially explain the differences between the observations and our model.

The CN emission is moderately optically thick in some of the hyper-fine structure lines, especially in the strongest blended lines.
The best fitting model roughly follows the shape of the radial distribution (see Fig.~\ref{fig:gas_radprofs}), but the modeled total flux is almost a factor of two too high.
Such a difference is entirely consistent with uncertainties in chemical reaction rate coefficients of the outer disk nitrogen chemistry.
Changing the nitrogen abundance in the outer disk does not have a large effect on the CN abundance (see Fig.~\ref{fig:paradependence}) and there is no clear physical reason to assume that the nitrogen abundance is lower than the ISM value \citep[see also][]{2015PNAS..112.8965B}.
A low gas mass is able to reproduce the CN line, as we show in Fig.~\ref{fig:gas_radprofs}. 
Other important parameters have hardly any effect due to the high elevation of the emitting layer in the disk (see App. \ref{app:paradependence}).
The CN emission is located mainly between a height of $z/r$~=~0.2-0.3, higher than the other tracers analyzed in this work.

\subsection{Disentangling gas mass and elemental abundances}
The gas mass and elemental abundances are interdependent for most tracers. 
For optically thin CO isotopologues, for example, one determines the total CO mass rather than either the CO abundance or gas mass.
To break this degeneracy we ran a grid of 60 models with varying mass and carbon abundance, as described in Sect. \ref{sssec:gasdistribution_fit}.
In Fig.~\ref{fig:gridplot} we present the integrated line flux of the key molecules for these models with disk gas mass ranging from 0.1~M$_\odot$ - 5$\times10^{-4}$~M$_\odot$ and system wide volatile carbon depletion between 1 and 32 with respect to the ISM.
We overplot the observed total line fluxes as a dashed white line and colored contours at a factor of 2 for comparison.
The colored regions are combined in the main panel for easy comparison of the molecules.

\subsubsection{HD}
The HD flux is independent of the carbon abundance, as expected, and changes almost directly proportional with the disk gas mass.
This indicates that changes in temperature corresponding to the variations in the gas mass do not have strong influence on the emitting region of HD.
The observed upper limit of HD constrains the gas mass to be lower than 0.05 M$_\odot$, which gives reliable evidence that the total disk gas mass is lower than expected from the total dust mass (5.8$\times10^{-4}M_{\odot}$) and the canonical ISM value for the gas-to-dust ratio of 100.

\subsubsection{CO isotopologues}
\ce{C^{17}O} has a near linear relationship between carbon abundance and gas mass, which indicates that the emission is fully determined by the total CO mass in the system.
The other CO isotopologue emission lines follow similar trends as \ce{C^17O}, but are slightly less sensitive to changes in gas mass, as these lines are optically thick at least in some parts of the disk.
The optically thick \ce{^12CO} line prefers lower gas masses, for which the emitting surface moves to colder layers deeper inside the disk, decreasing the modeled flux at larger radii (see Sect. \ref{ssec:results gasdistribution}).
The total flux of \ce{^12CO} changes by only a factor of a few over the two orders of magnitudes in gas mass, which shows that the temperature dependence of the optically thick tracers is much less sensitive to the gas mass than the density dependence of the optically thin tracers.

\subsubsection{N-bearing species}
We find that the \ce{N_2H+} emission is strongly dependent on the total gas mass in the system, changing almost in direct proportion with it.
\ce{N_2H+} behaves opposite to CO: it increases in line strength by lower carbon abundances due to less competition with CO for \ce{H$_3$+}, and less conversion from \ce{N2H+} to \ce{HCO+} by direct reaction with CO.
The CN flux varies by only a factor of $\sim$3 over two orders of magnitude change in gas mass, and is roughly independent of the carbon abundance.

\subsubsection{Combining tracers}
Without a robust detection of HD, it is not possible to determine either the C/H abundance or the gas mass because all known other tracers vary as function of both quantities.
The optically thick tracers CO and CN could in principle be used, but their dependence on the temperature is weak, and the temperature structure is less constrained in the models than the density distribution.
However, combining the different tracers (Fig.~\ref{fig:gridplot}) we find a confined region in the modeling parameter space, $M_\mathrm{g}=0.01 ^{+0.01}_{-0.004}~M_{\odot}$; C/H~=~$3 \pm 1.5 \times10^{-5}$, that agrees within a factor of 2 with the CO isotopologue fluxes, HD upper limit, CN and \ce{N_2H+}.
All optically thin tracers are consistent within the observational uncertainty, assuming the usual 10\% calibration error.

\subsection{Determination of the C/O ratio}
The \ce{C_2H} $N$=3-2 transition changes as function of carbon abundance, but is also prone to variations in the C/O ratio. 
For that reason we leave this species out of the analysis for the carbon abundance in Fig.~\ref{fig:gridplot}. 
However, we can use \ce{C_2H} to deduce the C/O ratio in the system and determine the O/H abundance.
We find that the integrated line flux of the \ce{C_2H} $N$=3-2 transition in our model is not very sensitive to the C/O ratio (see Fig.~\ref{fig:gas_radprofs}) compared to other sources \citep[see e.g.,][]{2021ApJS..257....7B}.
A change in C/O ratio from ISM level ($\sim$0.47) to unity changes the total \ce{C_2H} emission by only a factor of 2, but a little bit of excess carbon, not locked up in CO, (i.e., C/O ratio of 1.1) would make a difference (see Fig.~\ref{fig:gas_radprofs}) since this changes the carbon chemistry considerably. 
The bulk of the \ce{C_2H} emission in our model originates from a region at 250-750 AU, that is contributing less prominently in the observed azimuthally averaged radial profile.
The maximum resolvable scale of $r \sim600$ AU for that data set should be large enough to pick up this emission in the disk.
Changing the C/O ratio in the model mainly has an impact on the emission inside 200 AU.
With ISM like C/O ratio we under-produce the \ce{C_2H} emission at $r<200$~AU with an order of magnitude (see Fig.~\ref{fig:gas_radprofs}).
To determine the C/O ratio we ran a grid of models varying the C/O ratio by depleting more oxygen in a range between 0.47 and 1.5.
This procedure is repeated for a more massive model (green line in Fig. \ref{fig:gas_radprofs}).
An elevated C/O ratio $\sim$ 1 matches consistently the centrally peaked emission component, which means that the volatile elemental oxygen abundance in the best fitting model, 3.4$\times 10^{-5}$, is depleted by a factor of 8 with respect to the ISM in the best fitting model.

\section{Discussion}
\label{sec:discussion}
\subsection{Comparison with literature gas masses}
Previously, the gas mass of LkCa~15 has been inferred from dust continuum observations \citep{2012ApJ...747..136I}, optically thick \ce{^12CO} and \ce{^{13}CO} emission \citep{2019ApJ...881..108J} and the HD upper limit \citep{2016ApJ...831..167M}.
We find with our method a gas mass of $M_\mathrm{g}=0.01 ^{+0.01}_{-0.004}~M_{\odot}$.
This value agrees with the high-resolution continuum observations and observed dust mass, but the total mass of the system is lower than values usually assumed in the literature due to the lower gas-to-dust ratio.
We find that the gas mass of LkCa~15 is an order of magnitude lower than the value previously reported in \citet{2019ApJ...881..108J}.

There are three key differences between the models described in \citet{2019ApJ...881..108J} and those presented in this work.
The first is the assumed density structure in radial and vertical direction. 
Our models follow the standard formulation for viscous disks with a power law and exponential taper (see Eq. \ref{eq:surface_density_profile}) and a gaussian vertical distribution (see Eq. \ref{eq:small_grains_distribution}). 
\citet{2019ApJ...881..108J} neglect the exponential taper and solve the hydrostatic equation to find a self-consistent vertical density structure.
Neglecting the exponential taper leads to very high values of the power law index; they find $\gamma~=~4$, compared to our $\gamma~=~1.5$.
They show that using lower values for $\gamma$, which reproduces their \ce{^13CO} emission but not \ce{^12CO}, leads to values for the gas mass that are consistent with that found in this work.

The second difference is the temperature structure. 
\citet{2019ApJ...881..108J} mainly analyze the optically thick \ce{^12CO} 3-2 emission that traces the temperature structure rather than the density structure in the disk, which is much less sensitive to the mass than optically thin tracers (see Fig.~\ref{fig:gridplot}).
Furthermore, their temperature structure is calculated only by the dust grains and the hydrostatic equation, neglecting any gas heating by the photo-electric effect or cooling by collisions.
This leads to an nonphysical, almost vertically isothermal temperature structure and a midplane CO snowline at $~300$~AU that is inconsistent with observations described in Sect. \ref{ssec:results gasdistribution} that put the CO snowline at $~60$~AU.

The third difference is the chemistry that is included. \citet{2019ApJ...881..108J} used a parametrized CO abundance structure assuming a CO abundance of 1.4$\times10^{-4}$, including freezeout and photo-dissociation, but neglecting any CO chemistry, self-shielding and depletion.
With this chemical approach and the adopted gas temperature structure they need to vary the \ce{^12C}/\ce{^13C} ratio radially by almost four orders of magnitude to reproduce both the \ce{^12CO} and \ce{^13CO} emission profiles.
The strength of this work is that we reproduce self-consistently the total flux and radial emission profiles of optically thick CO, many optically thin CO isotopologue lines, \ce{C_2H}, \ce{N2H+} and take the HD upper limit into account.

\subsection{Comparison of HD with observations}
The mass upper limit from HD that we find in this work is more constraining than previous models by \citet{2016ApJ...831..167M}.
With new evidence from the interferometric data analyzed in this paper, we use a more extended disk structure with a lower average gas temperature.
High angular resolution continuum observations have additionally moved the continuum cavity radius further outwards from 39 AU to 63 AU, which has significant effect on the heating in the region where most HD emission originates ($<$250 AU).
The chemical modeling of HD is determined self-consistently with the other chemistry and we include the continuum rings that are important for the temperature balance in the HD emitting region.
We show that the present \textit{Herschel} upper limit to HD is less constraining for disentangling the gas mass from the elemental C and O abundances than the CO isotopologues and \ce{N_2H^+}. 
However, a robust detection, which would be constraining, would only require a sensitivity ten times better than that of the Herschel observation that yielded the upper limit used here \citep{2016ApJ...831..167M}.
A 3$\sigma$ detection would require a channel RMS of 6.3$\times10^{-19}$~W m$^{-2}$ $\mu$m$^{-1}$.

\subsection{Modeling of \ce{N_2H+}}
\label{ssec:modelingN2h+}
\ce{N_2H+} is often raised as a potential mass and CO abundance degeneracy breaker, because of its inverse trend with the CO abundance. 
We retrieve similar trends for LkCa~15 as described in \citet{2019ApJ...881..127A} and confirmed by \citet{2022arXiv220109900T} and \citet{2022arXiv220200709A}.
High \ce{N_2H+} fluxes are found for models with a high gas mass and low carbon abundance, though the observed \ce{N_2H+} flux in this particular case is consistent with moderate depletion and a low gas mass.

The largest uncertainties on the \ce{N_2H+} modeling are the total ionization and the volatile nitrogen abundance.
In our modeling we assume that the ionization is dominated by cosmic rays.
The \ce{N2H+} flux over two observed epochs is similar \citep{2019ApJ...882..160Q,2020ApJ...893..101L}, which is consistent with ionization set primarily by cosmic rays rather than by (variable) X-rays, which can lead to variations in \ce{HCO+} and \ce{N2H+} strength \citep{2017ApJ...843L...3C}.
The value for $\zeta$ assumed in this work is on the high end for protoplanetary disks. 
Running a small grid of models with different values for $\zeta$ we find that one order of magnitude change in $\zeta$ leads to a change in \ce{N2H+} flux by a factor of three (see App. \ref{app:paradependence}), which is consistent with the theoretical $\sqrt{\zeta}$ relation.
Our assumed uncertainty on the modeling is consistent with CR ionization rates between $10^{-16}-10^{-18}$ s$^{-1}$.
The minimum CR ionization rate that is still consistent with the HD upper limit is  $10^{-18}$ s$^{-1}$, assuming ISM N/H abundance.
Future observations of \ce{HCO+} will help to determine the ionization level in the disk and decrease the uncertainty on the disk gas mass.

As explained in Sect. \ref{sec:intro}, nitrogen is thought to be less affected by volatile depletion processes than carbon and oxygen.
Unfortunately, there are no other nitrogen carrying molecules that could constrain the nitrogen abundance and give a handle on the uncertainty of the abundance in the disk. 
However, using the strict upper limit on the gas mass for the HD observations, we can determine lower limits on the nitrogen abundance based on the \ce{N2H+} flux, which is sensitive to the nitrogen abundance (see also App. \ref{app:paradependence}).
Varying the nitrogen abundance between 100 - 0.1 ppm for CR ionization rates 5$\times10^{-17}$ and $\times10^{-18}$s$^{-1}$ gives strict lower limits of 3.8 and 17 ppm on the N/H ratio, respectively, using the gas mass upper limit from the HD observations and corresponding carbon abundance.

\subsection{Scientific implications}
\subsubsection{Gas-to-dust ratio}
Our inferred gas mass for the LkCa~15 disk from thorough modeling of many molecular transitions, including \ce{C^17O} 2-1, \ce{N2H+} 3-2 and the HD 1-0 upper limit is $M_\mathrm{g}=0.01 ^{+0.01}_{-0.004}~M_{\odot}$.
These values correspond to a midplane gas-to-dust ratio of 10-50 with locally very low gas-to-dust ratios $<10$ as a result of the two bright continuum rings attributed with dust traps.
Low gas-to-dust ratios in the two continuum rings are consistent with stability related arguments \citep{2020A&A...639A.121F}.
The large amount of dust and the vicinity of the midplane CO snowline just outside the cavity at 63 AU (see Fig.~\ref{fig:dali_overview}) makes this region very suitable for triggering streaming instabilities
\citep[see e.g.,][]{2010ApJ...722L.220B}.

\subsubsection{C/H ratio}
Multiple studies find observational evidence for an increase in volatile depletion (on top of the standard freeze-out and photo-dissociation processes) with the age of the system \citep{2020ApJ...898...97B,2020ApJ...891L..17Z,2022arXiv220104089S}. 
The moderate level of carbon depletion, by a factor $f_\mathrm{dep}~=~3-9$ with respect to the ISM, that we find for LkCa 15, follows this trend given the young age of the system \citep[2 (0.9-4.3) Myr][]{2020ApJ...890..142P}.
Younger stars typically have higher accretion rates, emit stronger in ultra-violet wavelengths and therefore have warmer disks than older sources. 
The disk temperature structure is thought to play an important role in setting the CO abundance and C/O ratio in the disk, as it determines the size of the region in the disk where CO can freeze out \citep{2021A&A...653L...9V}. 

However, LkCa~15 could be considered a typical ``cold transition disk", with a large dust mass in regions where CO could freeze out and at least one dust trap beyond the CO snowline that potentially halts the radial drift of CO rich ice.
This implies that the level of CO depletion in cold disks might be the result of accumulative conversion of CO on longer timescales, resulting in the evolutionary trend seen in the observations of T Tauri stars.
This is in line with modeling of CO conversion and locking up in pebbles that typically happen at timescales of $\sim10^5-10^6$ year \citep[see e.g.,][]{2020ApJ...899..134K}.

\subsubsection{C/O ratio}
The observed \ce{C_2H} flux in LkCa~15 is consistent with a C/O ratio of $\sim$1.
This value is in agreement with the O/H ratio inferred from deep \textit{Herschel} observations that, unexpectedly, do not find any water emission in LkCa~15 \citep{2017ApJ...842...98D}.
The water could be hidden if it is only frozen out in parts of the disk with millimeter emission, but likely water is depleted by a factor of at least 5-10 throughout the whole disk, similar to the explanation of the observed flux in HD 100546 and the upper limits of the full sample of Water In Star-forming regions with \textit{Herschel} (WISH) sample, resulting in elevated C/O ratios \citep{2017ApJ...842...98D, 2021A&A...648A..24V}, \citep[see also][]{2022arXiv220710744P}.
Comparison with rotational water lines would be very interesting, but was not possible in this case, as there are no observations in HDO and the disk is too cold for observations of the \ce{H_2^{18}O} line at 203 GHz. 
The C/O ratio of 1 that we find for LkCa 15 is low compared to other sources like TW Hya C/O~=~1.5-2 \citep{2016A&A...592A..83K}. A possible limitation of the C/O elevation is efficient vertical mixing stirring up oxygen rich ice to warmer regions in the disk where it releases oxygen to the gas.

\citet{2019ApJ...876...25B} find that either high UV penetration or a high level of CO depletion is necessary inside 200 AU to explain the centrally peaked \ce{C_2H} emission component, and low level CO depletion beyond 200 AU to explain the extended emission. 
We show that a constant level of carbon depletion $f_\mathrm{dep}~=~3-9$ reproduces both features in our model.
A combination of the very flat geometry and deep inner gas cavity enables deep UV penetration. 
Together with an elevated C/O ratio this produces enough \ce{C_2H} flux in the inner regions of the disk, without the need for additional carbon depletion.

\section{Conclusions}
\label{sec:conclusions}
Determining the gas mass and volatile elemental abundances of C and O in protoplanetary disks is crucial in our understanding of planet formation.
In this work we present new NOEMA observations of LkCa~15 in the CO isotopologues \ce{C^17O}, \ce{C^18O}, \ce{^13CO} and CN. Combining these observations with archival \ce{N2H+}, \ce{C_2H} and HD data and high resolution CO isotopologue data we constrain the gas mass and C/H and O/H abundances in the disk using physical-chemical modeling.
We summarize our findings below:
\begin{itemize}
    \item Using optically thick \ce{CO} and optically thin CO isotopologue lines we are able to construct a model that reproduces all analyzed disk integrated emission fluxes within a factor of two. Radial profiles of continuum and emission lines are reproduced by the model in all cases.
    \item Using the combination of \ce{N2H+}, \ce{C^17O} and the HD upper limit we constrain the gas mass of LkCa~15 to be $M_\mathrm{g}=0.01 ^{+0.01}_{-0.004}~M_{\odot}$ compared with $M_\mathrm{d}=5.8\times10^{-4}~M_{\odot}$.
    This is consistent with cosmic ray ionization rates between $10^{-16} - 10^{-18}$ s$^{-1}$, where $10^{-18}$ s$^{-1}$ is a lower limit based on the HD upper limit.
    This gas mass implies that the average gas-to-dust ratio in the system is lower than the canonical value of 100, which means that this particular system could be efficient in producing planetesimals via the streaming instability, especially at the location of the dust traps.
    \item The CO gas abundance relative to \ce{H2} is only moderately reduced compared with the ISM in the LkCa~15 disk by a factor between 3-9. This reduced depletion of elemental carbon is consistent with the age of the system, but contrast with the higher depletions seen in older cold transition disks. This contrast suggests that the long timescales for CO transformation and locking up in pebbles contribute to the variation in the level of carbon depletion seen in T Tauri stars at different stages in their evolution.
    \item The \ce{C_2H} emission in the LkCa~15 disk is consistent with a C/O ratio around unity in the bulk of the disk. These findings agree with the non-detection of water in deep $Herschel$ observations. Given that the LkCa~15 disk is one of the brightest sources in \ce{C_2H}, this proves that not all sources with high \ce{C_2H} flux require C/O ratios significantly higher than unity.
\end{itemize}
We find that a combination of CO isotopologues, N-bearing species and \ce{C_2H} provides good constraints on protoplanetary disk masses, the volatile C/H abundance and the C/O ratio. Far-infrared HD detections, and complimentary ionization tracers like \ce{HCO+} would greatly strengthen the mass constraints.

\begin{acknowledgements}
This paper makes use of the following ALMA data:
2018.1.01255.S 
2018.1.00945.S
2017.1.00727.S
2016.1.00627.S
2015.1.00657.S
ALMA is a partner-ship of ESO (representing its member states), NSF (USA), and NINS (Japan), together with NRC (Canada) and NSC and ASIAA (Taiwan), in cooperation with the Republic of Chile. The Joint ALMA Observatory is operated by ESO, AUI/NRAO, and NAOJ.
This paper is based on observations carried out under project number S20AT with the IRAM Interferometer NOEMA. IRAM is supported by INSU/CNRS (France), MPG (Germany) and IGN (Spain).
M.L. acknowledges support from the Dutch Research Council (NWO) grant 618.000.001.
Astrochemistry in Leiden is supported by the Netherlands Research School for Astronomy (NOVA), by funding from the European Research Council (ERC) under the European Union’s Horizon 2020 research and innovation programme (grant agreement No. 101019751 MOLDISK).
We thank the IRAM staff member Orsolya Feher for assistance of observations and data calibrations.
\end{acknowledgements}

\bibliographystyle{aa}
\bibliography{references.bib}

\begin{appendix}
\section{\ce{HC_3N} detection}
\begin{figure*}[!b]
    \centering
    \includegraphics[width=0.75\textwidth]{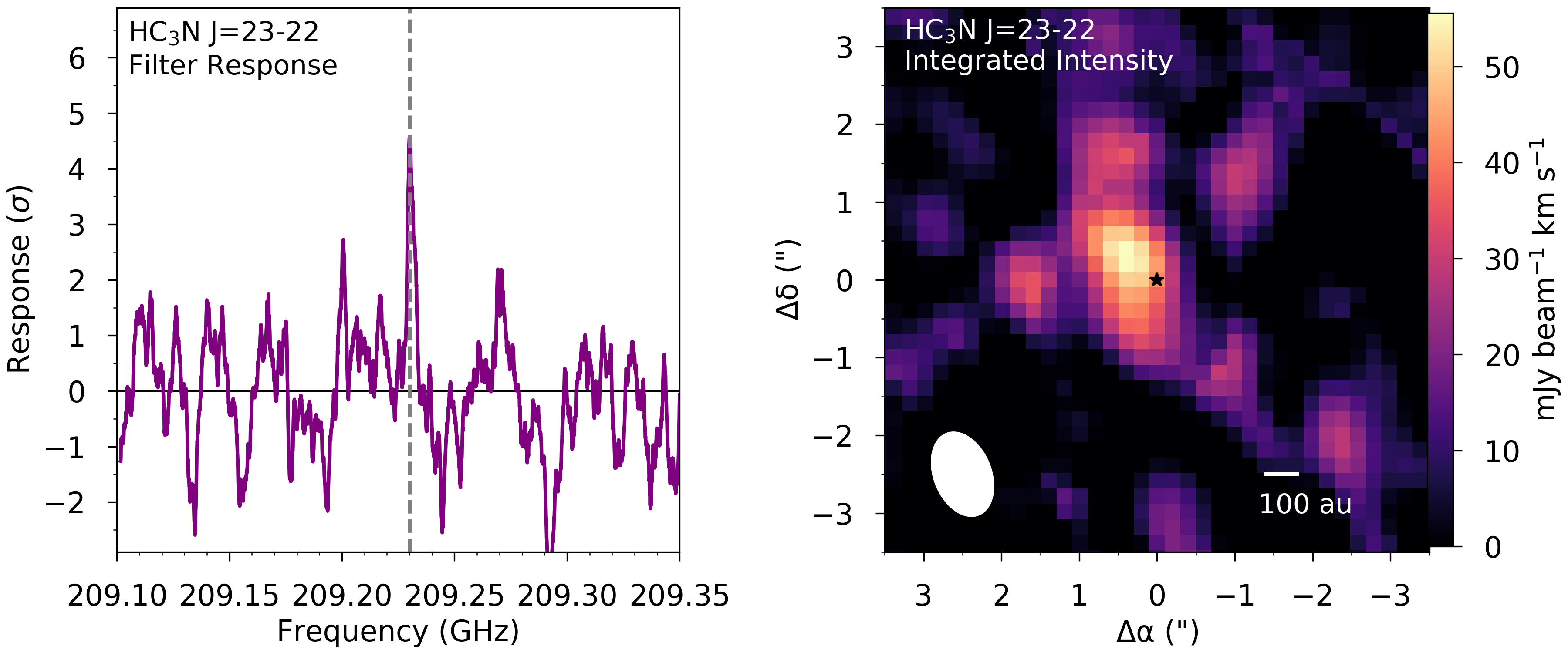}
    \caption{Left: filter response for a Keplerian mask, the \ce{HC3N} $J$=23-22 transition is marked with a dashed line. Right: Keplerian masked integrated intensity map. The beam is shown in the bottom left corner.}
    \label{fig:hc3n}
\end{figure*}

We also report the detection of \ce{HC3N} in the LkCa 15 disk. 
We identified the \ce{HC3N} $J$=23-22 transition at a rest frequency of 209.230234 GHz with an upper energy level of 120.5K \citep{2016JMoSp.327...95E} using matched filter analysis \citep[see][]{2018AJ....155..182L}. 
Using a Keplerian mask set by the source properties given in Table \ref{tab:source_props}, the response reached a significance of 4.6 at the correct frequency for the \ce{HC3N} J=23-22 transition. 
The stacked \ce{HC3N} J=31-30, J=32-31 and J=33-32 transitions were reported as a potential marginal detection in \citep{2020ApJ...893..101L}. 
In Fig.~\ref{fig:hc3n} we show the impulse response spectrum with a 100 AU Keplerian mask which reaches a peak of 4.6 $\sigma$. 
The line is imaged using natural weighting, and at a channel width of 1 km s$^{-1}$ the line reached a peak of 26.8 mJy/beam with a S/N ratio of 4.5.
In Fig.~\ref{fig:hc3n} we show the Keplerian masked integrated intensity map which highlights that the emission is compact and mostly within one beam.
The integrated flux of the line is 0.18~Jy~km~s$^{-1}$, which results in an average column density of 1-5$\times10^{13}$~cm$^{-2}$ assuming a range of excitation temperatures between 30-60 K \citep[similar to that found in ][]{2021ApJS..257....9I} and an emitting area of 1 beam.
This number is consistent with disk-integrated column densities found in the large program Molecules with ALMA at Planet-forming Scales (MAPS), where they find column densities ranging from 2-8 $\times10^{13}$~cm$^{-2}$ in a range of rotational temperatures between 30-60 K in the sources MWC 480, HD 163296, AS 209 and GM AUR.
The abundance of \ce{HC3N} in disk models has been shown to vary with C/O ratio \citep{2019ApJ...886...86L}.
Therefore, the detection of \ce{HC3N} in the LkCa 15 disk is consistent with the strong \ce{C_2H} line fluxes and elevated C/O ratio with respect to the ISM. 

\section{Modeling robustness}
\label{app:paradependence}
The effect of the C/O ratio, C/H abundance and the gas mass on the model is discussed in detail in Sect. \ref{sec:results}. In this section the effect of other parameters such as the N/H abundance, $\zeta$, $f_\mathrm{\ell}$ and the extra dust in the cavity is investigated.

\subsection{parameter dependence}
The results for a grid with varying specific model parameters are presented in Fig.~\ref{fig:paradependence}. 
We vary the nitrogen abundance between $10^{-4} - 10^{-7}$, $\zeta$ between $10^{-16} - 10^{-19}$~s$^{-1}$ and $f_\mathrm{\ell}$ between 0.99 and 0.5.
Variations in the nitrogen abundance have no effect on the CO lines as these abundances are constrained by the availability of oxygen for C/O = 1.
CN depends moderately on the nitrogen abundance, changing only a factor of two over 4 orders of magnitude difference in C/N.
This means that CN is not a dominant carrier of N but that its abundance is set by other quantities such as the UV flux.
The \ce{N2H+} emission is very sensitive to the nitrogen abundance, as the CO/\ce{N_2} ratio has to be low to form \ce{N2H+} efficiently (see the discussion in Sect. \ref{ssec:gasdistr_nbearing_species}).
Unfortunately, we do not have the tools to determine the N/H abundance separately, considering CN is not sensitive to the N/H abundance. 
Nitrogen is thought to be less depleted than carbon \citep{2015PNAS..112.8965B,2018ApJ...865..155C, 2018A&A...615A..75V}, given that the level of carbon depletion is low in this system the assumption of ISM level N/H is valid, but requires additional attention in the future.

Changing $\zeta$ has only effect on the optically thin CO lines as the rare CO isotopologues are less efficient in self-shielding against destruction.
The ionization balance in the disk is mainly set by $\zeta$, which results in a moderate dependence of \ce{N2H+} on $\zeta$ as discussed in Sect. \ref{ssec:modelingN2h+}.

Lowering $f_\mathrm{\ell}$ will impact the vertical temperature structure as more surface area per dust mass unit is moved to the upper layers of the disk. 
High values for the large grain fractions are motivated by studies that find that mm-size dust can form on timescales of 10$^5$ yr \citep{2012MNRAS.425.3137U,2016SSRv..205...41B}.
Lower values of $f_\mathrm{\ell}$ are one of the only options to decrease the CN and \ce{^12CO} flux and would increase the mass upper limit of the HD emission as it decreases the size of the HD emitting region.
However, the effect on CN and \ce{^12CO} is small and physical values <$f_\mathrm{\ell}$ = 0.9 result in underproducing the CO isotopologue emission even taking higher C/H abundances and gas masses into account.

\subsection{dust component in cavity}
\begin{figure*}
    \centering
    \includegraphics[height=0.9\textheight]{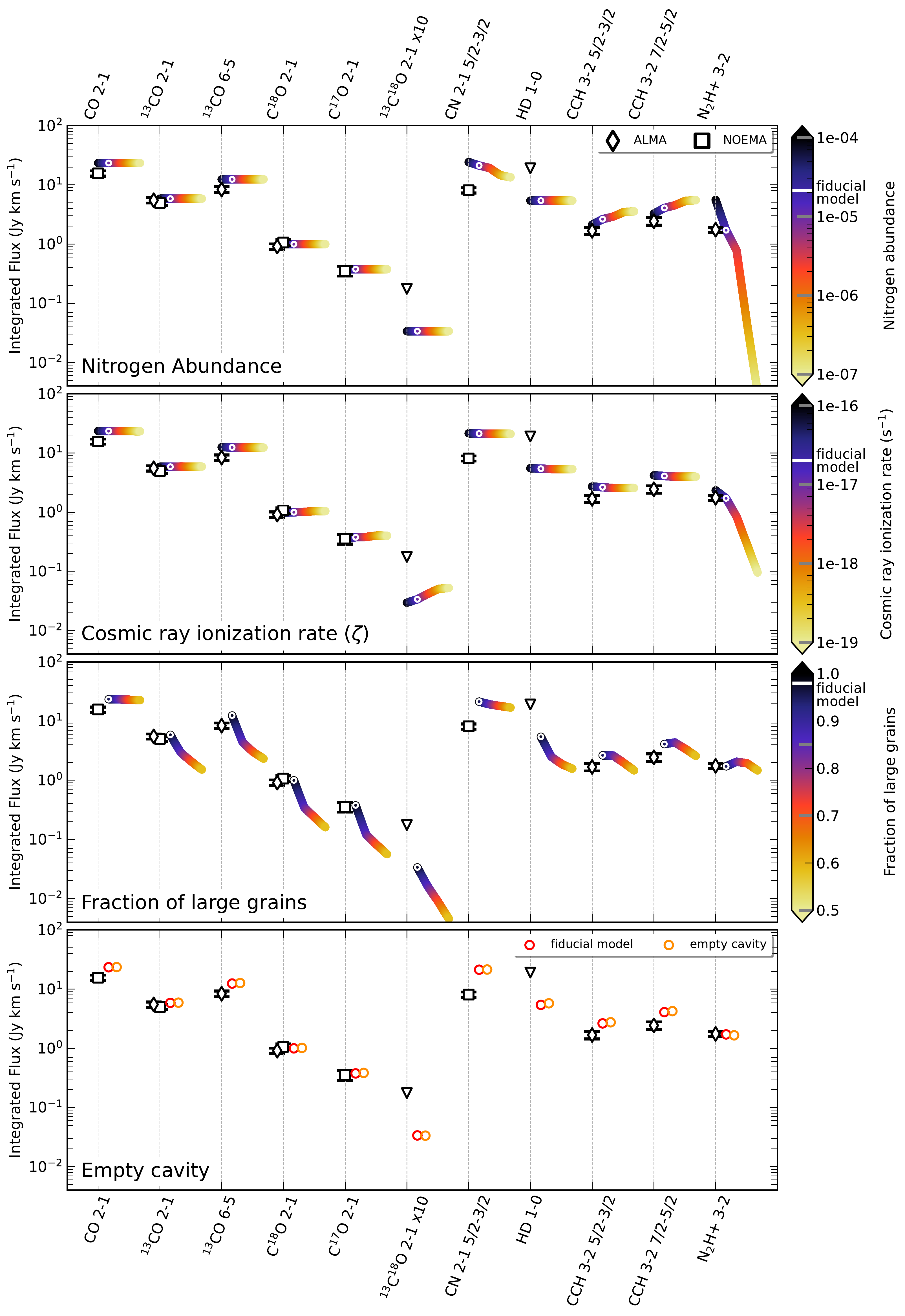}
    \caption{Total flux dependencies of the various molecules on the main parameters in the model. From top to bottom: Nitrogen abundance, cosmic ray ionization rate $\zeta$, fraction of settled large grains $f_\mathrm{\ell}$. Data grid points and the fiducial model are shown in the colorbar as grey and white tickmarks, respectively. The color bars represent interpolated values between these models. The fiducial model is shown in the plots as a white dot. The bottom panel shows the result for a similar model as the fiducial model, but with a cleared cavity between 1-63 AU (see Fig.~\ref{fig:model_dust_setup}). Note that the fiducial model is the same as the "best" model in Fig.~\ref{fig:gas_radprofs}.}
    \label{fig:paradependence}
\end{figure*}

Motivated by the millimeter dust continuum observations presented in \citet{2020A&A...639A.121F} we included a small contribution of dust inside the cavity to improve the fit on the millimeter continuum radial profile (see Fig.~\ref{fig:model_dust_setup}).
This third dust ring could be a result of diffusing small grains inside the cavity or a planet. 
In the bottom panel of Fig.~\ref{fig:paradependence} we compare the model fluxes with a model that uses the conventional transition disk setup with a cleared cavity and a small 1 AU inner disk (following the dashed line in Fig.~\ref{fig:model_dust_setup}).
The dust surface density in the cavity in our fiducial model does not have a significant effect on the disk integrated line fluxes.

\end{appendix}
\end{document}